\documentstyle[12pt,a4,epsfig,cite]{article}
\newcommand{\beq}{\begin{equation}}
\newcommand{\eeq}{\end{equation}}
\newcommand{\beqa}{\begin{eqnarray}}
\newcommand{\eeqa}{\end{eqnarray}}
\newcommand{\beqan}{\begin{eqnarray*}}
\newcommand{\eeqan}{\end{eqnarray*}}
\newcommand{\no}{\nonumber}

\newcommand{\letsmod}{\stackrel{<}{_\sim}}

\newcommand{\ol}{\overline}
\newcommand{\ra}{\rightarrow}

\newcommand{\ve}{\varepsilon}
\newcommand{\vp}{\varphi}

\newcommand{\dg}{\dagger}
\newcommand{\wt}{\widetilde}
\newcommand{\wh}{\widehat}
\newcommand{\dfrac}{\displaystyle \frac}

\newcommand{\ben}{\begin{enumerate}}
\newcommand{\een}{\end{enumerate}}
\newcommand{\bfl}{\begin{flushleft}}
\newcommand{\efl}{\end{flushleft}}
\newcommand{\ba}{\begin{array}}
\newcommand{\ea}{\end{array}}
\newcommand{\btab}{\begin{tabular}}
\newcommand{\etab}{\end{tabular}}
\newcommand{\bit}{\begin{itemize}}
\newcommand{\eit}{\end{itemize}}

\newcommand{\Ha}{{\cal H}}

\newcommand{\cL}{{\cal L}}
\newcommand{\M}{{\cal M}}
\normalsize
\begin{document}
\parskip=4pt plus 1pt  
\begin{titlepage}
\begin{flushright}
CERN-TH.6920/93 \\
UWThPh-1993-22
\end{flushright}
\vspace{1.5cm}
\begin{center}
{\Large \bf  Non-Leptonic Kaon Decays and the Chiral Anomaly*}\\[40pt]
{\bf G. Ecker, H. Neufeld } \\
Institut f\"ur Theoretische Physik \\
Universit\"at Wien \\
Boltzmanngasse 5, A-1090 Wien \\[5pt]
and \\[5pt]
{\bf A. Pich} \\
CERN \\
CH-1211 Geneva 23
\vfill
{\bf Abstract} \\
\end{center}
\noindent
An investigation is performed of all non-leptonic kaon decays sensitive to
the chiral anomaly. Within the framework of chiral
perturbation theory, there are two classes of anomalous amplitudes
at $O(p^4)$:
reducible and direct contributions. Only radiative transitions
are affected by the anomaly. The phenomenology of the decays
$K_L \to \pi^+ \pi^- \gamma$ and $K^+ \to \pi^+ \pi^0 \gamma$ is
studied in detail. Including the dominant contributions of $O(p^6)$,
the experimentally observed dependence of the direct emission amplitude
for $K_L \to \pi^+ \pi^- \gamma$ on the photon energy can be understood.
A survey is made of the rare ``anomalous'' decays $K \to \pi \pi \gamma
\gamma$ and $K \to 3 \pi \gamma (\gamma)$, including some numerical
estimates.

\vfill
\noindent * Work supported in part by Fonds zur F\"orderung der
wiss. Forschung (Austria), Project Nos. P08485-TEC, P09505-PHY
(EURODAPHNE Collaboration) and by CICYT (Spain), grant No. AEN90-0040
\begin{flushleft}
CERN-TH.6920/93 \\
June 1993
\end{flushleft}
\end{titlepage}
\renewcommand{\theequation}{\arabic{section}.\arabic{equation}}
\setcounter{equation}{0}
\section{Introduction}
The standard model is a chiral quantum field theory. The chiral
structure is responsible for the existence of the chiral anomaly
\cite{ABJ,Bard} whose theoretical origin and mathematical properties are well
understood. On the other hand, experimental tests of this
important ingredient of modern particle physics are relatively rare.

The chiral anomaly manifests itself most directly in the low-energy
interactions of the pseudoscalar mesons
(some of the phenomenological aspects can be found
in the recent reviews \cite{Ksemi,EckMor,BijRev}). The appropriate
framework to study these effects is chiral perturbation theory
(CHPT) \cite{Wein79,GL1,GL2}. For the strong, electromagnetic and
semileptonic weak interactions, all anomalous Green functions
can be obtained from the Wess-Zumino-Witten (WZW) functional
\cite{WZW}. In contrast to most other aspects of the standard model
in the world of hadrons, the translation from the fundamental level
to the effective chiral level is unambiguous and free from
hadronization problems.

The chiral anomaly also appears in the non-leptonic weak interactions.
The purpose of this paper is to give a systematic account of all
non-leptonic $K$ decays where the anomaly contributes at leading order,
 $O(p^4)$. As already shown in previous work \cite{ENP1,BEP1}, only
radiative $K$ decays are sensitive to the chiral anomaly in the
non-leptonic sector. There are two different manifestations of the
anomaly: the reducible amplitudes \cite{ENP1}, which can again be derived
directly from the WZW functional, and direct contributions \cite{BEP1,Cheng},
which are subject to some theoretical uncertainties. We shall present
a self-contained and systematic discussion of those
contributions. As a special application, the two
most frequent ``anomalous'' decays $K_L \to \pi^+ \pi^- \gamma$ and
$K^+ \to \pi^+ \pi^0 \gamma$ will be analysed in detail,
including the dominant effects of $O(p^6)$. In both cases, the direct emission
amplitudes are dominated by the anomaly. A careful treatment of $O(p^6)$
effects is necessary to understand the experimentally observed
dependence of the direct emission amplitude on the photon energy
for the decay $K_L \to \pi^+ \pi^- \gamma$.

In Sect. 2, we summarize the chiral realization of the $\Delta S=1$
non-leptonic weak interactions in the meson sector. The relevant
terms of the  strong and weak Lagrangians of $O(p^4)$ are listed.
The distinction between reducible and direct anomalous amplitudes
is explained in Sect. 3. The direct weak anomaly functional is related
to the general weak Lagrangian of $O(p^4)$. A list of all non-leptonic
$K$ decays with local anomalous amplitudes of $O(p^4)$ is given. General
features of the decays $K \to \pi \pi \gamma$ are put together in
the following section. A general theorem on the structure of the
lowest-order amplitudes for those decays is formulated and discussed.
The dominant effects of $O(p^6)$ are expected to be due to vector
meson exchange.  The factorization model is proposed to estimate
the direct weak terms
related to V exchange.

We turn to the phenomenology of
$K \to \pi \pi \gamma$ decays in Sect. 5. For completeness, we
include a brief review
of the theoretical status of the decays $K_{L,S} \to \pi^0 \pi^0
\gamma$ and  $K_S \to \pi^+ \pi^- \gamma$ even though they are not
subject to anomalous contributions. Our main emphasis, however,
is on the anomalous decays $K_L \to \pi^+ \pi^- \gamma$ and
$K^+ \to \pi^+ \pi^0 \gamma$. In both cases, the bremsstrahlung
amplitudes are suppressed as are the one-loop amplitudes. A careful
analysis of the magnetic amplitudes of $O(p^6)$ is made since
experiments are already sensitive to those subleading contributions.
We suggest an interpretation of the measured slope parameter in the
$K_L$ decay \cite{BNL,E731,LV} and compare with previous work by
other authors. In Sect. 6, on overview of the remaining non-leptonic
$K$ decays sensitive to the anomaly is given. In addition to some
comments on the general structure of those decays, numerical
results are presented for two typical transitions, $K^+ \to \pi^+ \pi^0
\gamma \gamma$ and $K_L \to \pi^+ \pi^- \pi^0 \gamma$. Our findings
are summarized in Sect. 7. Two Appendices contain a proof of the
bremsstrahlung theorem of Sect. 4 and the definition of loop functions
encountered in Sect. 5.

\renewcommand{\theequation}{\arabic{section}.\arabic{equation}}
\setcounter{equation}{0}
\section{CHPT for non-leptonic weak interactions}
At low energies ($E \ll M_W$), the $\Delta S=1$ non-leptonic weak
interactions are described by an effective Hamiltonian \cite{GW}
\beq
\Ha_{\rm eff}^{\Delta S=1} = \frac{G_F}{\sqrt{2}} V_{ud} V^*_{us}
\sum_i C_i Q_i + {\rm h.c.} \label{Heff}
\eeq
in terms of Wilson coefficients $C_i$ and local four-quark operators
$Q_i$. The effective chiral Lagrangian for (\ref{Heff}) to lowest
order in the chiral expansion can be written as
($F$ is the pion decay constant in the chiral limit,
$F \simeq F_\pi = 93.2$~MeV):
\beq
\cL_2^{\Delta S=1} = G_8 F^4 \langle \lambda L_\mu L^\mu\rangle +
G_{27} F^4 \left( L_{\mu 23} L^\mu_{11} + \frac{2}{3} L_{\mu 21}
L^\mu_{13}\right) + {\rm h.c.} \label{L2w}
\eeq
$$
\lambda = \frac{1}{2} (\lambda_6 - i \lambda_7), \qquad
L_\mu = i U^\dg D_\mu U, \qquad
\langle A\rangle = \mbox{tr } A.
$$

The matrix field $U(\vp)$ incorporating the eight pseudoscalar Goldstone
boson fields transforms linearly under the chiral group $SU(3)_L \times
SU(3)_R$. The covariant derivative
\beq
D_\mu U = \partial_\mu U - i r_\mu U + i U \ell_\mu
\eeq
with $3 \times 3$ Hermitian matrix fields $\ell_\mu$, $r_\mu$ contains in
particular the photon field:
\beqa
r_\mu &=& v_\mu + a_\mu = e Q A_\mu + \ldots \no \\
\ell_\mu &=& v_\mu - a_\mu = e Q A_\mu + \ldots \\
Q &=& \frac{1}{3} \mbox{ diag }(2,-1,-1) \, , \no
\eeqa
where $Q$ is the quark charge matrix.

The coupling constants $G_8$, $G_{27}$ in (\ref{L2w}) measure the strength
of the two parts in the effective Hamiltonian (\ref{Heff}) transforming
as ($8_L,1_R$) and ($27_L,1_R$), respectively, under chiral rotations.
Neglecting the small $\Delta I = 1/2$ part of the 27-plet, the Lagrangian
(\ref{L2w}) produces the tree-level amplitudes
\beqa
A(K_1^0 \ra \pi^+ \pi^-) &=& 2iF(G_8 + G_{27}^{(3/2)})
(M_\pi^2 - M_K^2) \no \\
A(K^+ \ra \pi^+ \pi^0) &=& 3i F G_{27}^{(3/2)} (M_\pi^2 - M_K^2)
\label{Ktree} \\
G_{27}^{(3/2)} &=& \frac{5}{9} G_{27}. \no
\eeqa
Up to radiative and higher-order chiral corrections \cite{RafRing,KMW2},
the ratio
\beq
\frac{G_{27}^{(3/2)}}{G_8} = \frac{1}{32}
\eeq
is small (and positive), expressing the $\Delta I = 1/2$ rule in
$K \ra 2\pi$ decays, and $|G_8| \simeq 9 \cdot 10^{-6} \mbox{ GeV}^{-2}$.

At next-to-leading order in CHPT, the chiral Lagrangian $\cL_4^{\Delta S=1}$
is already quite involved \cite{KMW1}. We shall only need the octet
Lagrangian of $O(p^4)$. Employing the operator basis of Ref.
\cite{EKW1}, we write
\beq
\cL_4^{\Delta S=1} = G_8 F^2 \sum_i N_i W_i + {\rm h.c.}\label{L4w}
\eeq
with dimensionless coupling constants $N_i$ and octet operators $W_i$.
Referring to Ref. \cite{EKW1} for the complete Lagrangian, we list here
only those terms that will be needed in the following. To facilitate the
use of this effective Lagrangian, we write down the relevant operators
in two different representations commonly used in CHPT:
\beqa
W_{14} &=& i \langle \Delta \{ f_+^{\mu\nu}, u_\mu u_\nu\}\rangle =
i \langle \lambda \{ F_L^{\mu\nu} + U^\dg F_R^{\mu\nu} U, L_\mu L_\nu
\} \rangle \no \\
W_{15} &=& i \langle \Delta u_\mu f_+^{\mu\nu} u_\nu \rangle =
i \langle \lambda L_\mu (F_L^{\mu\nu} + U^\dg F_R^{\mu\nu} U) L_\nu
\rangle \no \\
W_{16} &=& i \langle \Delta \{ f_-^{\mu\nu}, u_\mu u_\nu\}\rangle =
i \langle \lambda \{ F_L^{\mu\nu} - U^\dg F_R^{\mu\nu} U, L_\mu L_\nu
\} \rangle \no \\
W_{17} &=& i \langle \Delta u_\mu f_-^{\mu\nu} u_\nu \rangle =
i \langle \lambda L_\mu (F_L^{\mu\nu} - U^\dg F_R^{\mu\nu} U) L_\nu
\rangle \no \\
W_{18} &=& \langle \Delta(f_{+\mu\nu} f_+^{\mu\nu} - f_{-\mu\nu}
f_-^{\mu\nu})\rangle = 2 \langle \lambda (F_L^{\mu\nu} U^\dg F_{R\mu\nu} U
+ U^\dg F_{R\mu\nu} U F_L^{\mu\nu})\rangle \label{Wi} \\
W_{28} &=& i \ve_{\mu\nu\rho\sigma} \langle \Delta u^\mu\rangle
\langle u^\nu u^\rho u^\sigma\rangle = i \ve_{\mu\nu\rho\sigma}
\langle \lambda L^\mu\rangle \langle L^\nu L^\rho L^\sigma\rangle \no \\
W_{29} &=& \langle \Delta [\wt f_+^{\mu\nu} - \wt f_-^{\mu\nu}, u_\mu
u_\nu]\rangle = 2 \langle \lambda [U^\dg \wt F_R^{\mu\nu} U,L_\mu
L_\nu]\rangle \no \\
W_{30} &=& \langle \Delta u_\mu\rangle \langle \wt f_+^{\mu\nu} u_\nu
\rangle = \langle \lambda L_\mu\rangle \langle (\wt F_L^{\mu\nu} +
U^\dg \wt F_R^{\mu\nu} U) L_\nu\rangle \no \\
W_{31} &=& \langle \Delta u_\mu\rangle \langle \wt f_-^{\mu\nu} u_\nu
\rangle = \langle \lambda L_\mu\rangle \langle \wt F_L^{\mu\nu} -
U^\dg \wt F_R^{\mu\nu} U) L_\nu\rangle \no
\eeqa
$$
U = u^2, \qquad \Delta = u \lambda u^\dg, \qquad
u_\mu = i u^\dg D_\mu U u^\dg = u L_\mu u^\dg , \qquad
f_\pm^{\mu\nu} = u F_L^{\mu\nu} u^\dg \pm u^\dg F_R^{\mu\nu} u.
$$
$F_L^{\mu\nu}$, $F_R^{\mu\nu}$ are the field strength tensors associated
to the external gauge fields $\ell_\mu$, $r_\mu$ \cite{GL2} and
$\wt F_{L,R\mu\nu} = \ve_{\mu\nu\rho\sigma} F_{L,R}^{\rho\sigma}$ are
their duals.

To calculate non-leptonic weak amplitudes, we also need the chiral
Lagrangian for the strong, electromagnetic and semileptonic weak
interactions. At lowest order, it is given by
\beq
\cL_2 = \frac{F^2}{4} \, \langle D_\mu U D^\mu U^\dg + 2B_0 \M
(U + U^\dg)\rangle \label{L2}
\eeq
in the notation of \cite{GL2}, where $\M$ is the diagonal quark mass
matrix and $B_0$ is related to the quark condensate.

Of the strong chiral Lagrangian of $O(p^4)$ \cite{GL2} we
shall encounter only one term,
\beq
\cL_4 = - i L_9 \, \langle F_L^{\mu\nu} D_\mu U^\dg D_\nu U +
F_R^{\mu\nu} D_\mu U D_\nu U^\dg\rangle + \ldots \label{L4}
\eeq
Like many of the weak couplings $N_i$ in (\ref{L4w}), the measurable
(renormalized) coupling constant $L_9^r(\mu)$ is scale-dependent due
to the divergences of the one-loop functional \cite{GL2}. In Sect.~5
we shall use the standard value \cite{GL2} $L_9^r(M_\rho) \simeq
6.9 \cdot 10^{-3}$.

Finally, a crucial ingredient of our analysis is the chiral
anomaly, which also enters the effective description at $O(p^4)$. It
will be dealt with in the following section.

\renewcommand{\theequation}{\arabic{section}.\arabic{equation}}
\setcounter{equation}{0}
\setcounter{section}{2}
\section{The chiral anomaly in the non-leptonic weak sector}
The contributions of the chiral anomaly to strong, electromagnetic and
semileptonic weak amplitudes can be expressed in terms of the
Wess-Zumino-Witten (WZW) functional \cite{WZW} $S[U,\ell,r]_{WZW}$.
If the vector currents are to be conserved, it has the following
explicit form :
\beqa
S[U,\ell,r]_{WZW} &=&-\dfrac{i N_C}{240 \pi^2}
\int d\sigma^{ijklm} \left\langle \Sigma^L_i
\Sigma^L_j \Sigma^L_k \Sigma^L_l \Sigma^L_m \right\rangle \label{WZW} \\*
 & & - \dfrac{i N_C}{48 \pi^2} \int d^4 x
\varepsilon_{\mu \nu \alpha \beta}\left( W (U,\ell,r)^{\mu \nu
\alpha \beta} - W ({\bf 1},\ell,r)^{\mu \nu \alpha \beta} \right)
\no \\
W (U,\ell,r)_{\mu \nu \alpha \beta} & = &
\left\langle U \ell_{\mu} \ell_{\nu} \ell_{\alpha}U^{\dg} r_{\beta}
+ \frac{1}{4} U \ell_{\mu} U^{\dg} r_{\nu} U \ell_\alpha U^{\dg} r_{\beta}
+ i U \partial_{\mu} \ell_{\nu} \ell_{\alpha} U^{\dg} r_{\beta}
\right.\no  \\
& & +~ i \partial_{\mu} r_{\nu} U \ell_{\alpha} U^{\dg} r_{\beta}
- i \Sigma^L_{\mu} \ell_{\nu} U^{\dg} r_{\alpha} U \ell_{\beta}
+ \Sigma^L_{\mu} U^{\dg} \partial_{\nu} r_{\alpha} U \ell_\beta
\no \\
& & -~ \Sigma^L_{\mu} \Sigma^L_{\nu} U^{\dg} r_{\alpha} U \ell_{\beta}
+ \Sigma^L_{\mu} \ell_{\nu} \partial_{\alpha} \ell_{\beta}
+ \Sigma^L_{\mu} \partial_{\nu} \ell_{\alpha} \ell_{\beta}  \\
& & -~ i\left.  \Sigma^L_{\mu} \ell_{\nu} \ell_{\alpha} \ell_{\beta}
+ \frac{1}{2} \Sigma^L_{\mu} \ell_{\nu} \Sigma^L_{\alpha} \ell_{\beta}
- i \Sigma^L_{\mu} \Sigma^L_{\nu} \Sigma^L_{\alpha} \ell_{\beta}
\right\rangle \no \\
& & -~ \left( L \leftrightarrow R \right) \no \eeqa
$$
\Sigma^L_\mu = U^{\dg} \partial_\mu U \qquad
\Sigma^R_\mu = U \partial_\mu U^{\dg} $$
$$ N_C = 3 \qquad \varepsilon_{0123} = 1  $$
where $\left( L \leftrightarrow R \right)$ stands for the interchange
$$
U \leftrightarrow U^\dg, \qquad \ell_\mu \leftrightarrow r_\mu,
\qquad \Sigma^L_\mu \leftrightarrow \Sigma^R_\mu . $$
The functional $S[U,\ell,r]_{WZW}$ conserves parity and it reproduces
the anomaly under chiral transformations in Bardeen's form \cite{Bard}.
The integration in the first term of Eq. (\ref{WZW}) is over a
five-dimensional manifold whose boundary is four-dimensional Minkowski
space. Since the integrand is a surface term, both the first and the
second term of $S_{WZW}$ are $O(p^4)$, according to the usual chiral
counting rules.

The chiral anomaly also contributes to non-leptonic weak amplitudes
starting at $O(p^4)$. We may distinguish between two different
manifestations of the anomaly.

\subsection*{i. Reducible anomalous amplitudes}
These amplitudes arise from the contraction of meson lines between a weak
$\Delta S = 1$ Green function and the WZW functional.
At $O(p^4)$, there can only be one such contraction and the weak vertex
must be due to the lowest-order non-leptonic Lagrangian
$\cL_2^{\Delta S=1}$ in Eq. (\ref{L2w}). The corresponding diagrams are
of the type shown in Fig. \ref{redaa}.
\begin{figure}
\epsfig{file=f1anom.eps,height=7.0cm}
\caption{Feynman diagram for a reducible non-leptonic anomalous amplitude.}
\label{redaa}
\end{figure}

Since $\cL_2^{\Delta S=1}$ contains bilinear terms in the meson fields,
the so-called pole contributions to anomalous non-leptonic amplitudes
can be given in closed form by a simultaneous diagonalization
\cite{EPR3} of the kinetic parts of the Lagrangians $\cL_2$ and
$\cL_2^{\Delta S=1}$. The corresponding local Lagrangian
(octet part only) is \cite{ENP1}:
\beq
\cL_{\rm an}^{\Delta S=1} = - \frac{ieG_8}{8\pi^2F} \wt F^{\mu\nu}
\partial_\mu \pi^0 K^+ \stackrel{\leftrightarrow}{D_\nu} \pi^- +
\frac{\alpha G_8}{6\pi F} \wt F^{\mu\nu} F_{\mu\nu}
\left(K^+ \pi^- \pi^0 - \frac{1}{\sqrt{2}} K^0 \pi^+ \pi^-\right) +
{\rm h.c.}
\label{Law}
\eeq
Here
$F_{\mu\nu} = \partial_\mu A_\nu - \partial_\nu A_\mu$ is the
electromagnetic field strength tensor, $\wt F_{\mu\nu} = \ve_{\mu\nu
\rho\sigma} F^{\rho\sigma}$ its dual and
$D_\mu \vp^\pm = (\partial_\mu \mp ieA_\mu)\vp^\pm$ denotes the
covariant derivative with respect to electromagnetism. In the limit of
CP conservation, the anomalous Lagrangian (\ref{Law}) contributes only
to the decays
\beq
K^+ \ra \pi^+ \pi^0 \gamma,\pi^+ \pi^0\gamma\gamma \qquad \mbox{and}
\qquad K_L \ra \pi^+ \pi^- \gamma\gamma
\eeq
with real or virtual photons.

There are of course other reducible anomalous amplitudes corresponding
to the diagram in Fig. \ref{redaa}. A generic example is provided by a
non-leptonic Green function where an external $\pi^0$ or $\eta$ makes
an anomalous transition to two photons. Such transitions are the
dominant $O(p^4)$ contributions to the decays $K_S \ra \pi^0\gamma\gamma$
\cite{EPR2} and $K_L \ra \pi^0\pi^0\gamma\gamma$ \cite{DFR,FK}.
All reducible anomalous amplitudes of $O(p^4)$ are proportional
to $G_8$ in the octet limit. No other unknown parameters are involved.

\subsection*{ii. Direct weak anomaly functional}
The second manifestation of the anomaly in non-leptonic weak amplitudes
arises diagrammatically from the contraction of the $W$ boson field
between a strong Green function on one side and the WZW functional on
the other side. However, such diagrams cannot be taken literally at a
typical hadronic scale, because of the presence of strongly interacting
fields on both sides of the $W$. Instead, one must as in Sect. 2 first
integrate out the $W$ together with the heavy quark fields. The
operators appearing in the operator product expansion must then be
realized at the bosonic level in the presence of the anomaly.

Following the methods of Ref. \cite{PR}, the bosonization of four-quark
operators in the odd-intrinsic parity sector was investigated in
Ref. \cite{BEP1}. As in the even-intrinsic parity sector, the bosonized
four-quark operators contain factorizable (leading in $1/N_C$, where
$N_C$ is the number of colours) and non-factorizable parts (non-leading
in $1/N_C$).

Due to the non-renormalization theorem \cite{AB} of the chiral anomaly,
the factorizable contribution of $O(p^4)$ can be calculated exactly
\cite{BEP1}. The bosonized form of a $(V-A) \times (V-A)$ four-quark
operator in the anomalous sector is [factorizable contribution of
$O(p^4)$]:
\beq
\ol{q_{lL}}\gamma^\mu q_{kL}\,\, \ol{q_{jL}}\gamma_\mu q_{iL}
\leftrightarrow \dfrac{\delta S_{WZW}}{\delta \ell^\mu_{lk}}\dfrac{\delta
S_2}{\delta \ell_{\mu,ji}} + \left(lk \leftrightarrow ji\right)
\label{fact} \eeq
where
\beqa
\dfrac{\delta S_2}{\delta \ell_{\mu,ji}} & = & - \dfrac{F^2}{2}
\left(L^\mu\right)_{ij} \\
L^\mu & = & i U^\dg D^\mu U       \label{curr2} \no
\eeqa
is the left-chiral current of lowest order $p$ corresponding to the
chiral Lagrangian (\ref{L2}).
The anomalous current [of $O(p^3)$] has the following form
\beqa
\dfrac{\delta S_{WZW}}{\delta \ell_{\mu,ji}} & = & \dfrac{1}{16 \pi^2}
\ve^{\mu \nu \alpha \beta} J^{an}_{\nu \alpha \beta,ij} \no \\*
J^{an}_{\nu \alpha \beta} & = & i L_\nu L_\alpha L_\beta
+ \left\{F^L_{\nu \alpha} + \frac{1}{2}U^\dg
F^R_{\nu \alpha} U, L_\beta \right\} . \label{Jan}
\eeqa
A physically irrelevant polynomial in the external fields $\ell,r$ has been
omitted in the anomalous current (\ref{Jan}).

Specializing to the dominant octet operator in $\Ha_
{\rm eff}^{\Delta S=1}$ [Eq. (\ref{Heff})],
\beqa
Q_- &=& Q_2 - Q_1 \\ \label{Q-}
Q_1 &=& \bar s \gamma^\mu (1-\gamma_5)d \,\, \bar u \gamma_\mu(1-\gamma_5)u
\no \\
Q_2 &=& \bar s \gamma^\mu (1-\gamma_5)u \,\, \bar u \gamma_\mu(1-\gamma_5)d,
\no
\eeqa
one obtains the following
bosonized form of $O(p^4)$ in the factorizable approximation for the
odd-parity part \cite{BEP1,Cheng}:
\beqa
Q_-(\mbox{fact}) & \leftrightarrow & \dfrac{F^2}{16 \pi^2}
\left(2 i \ve^{\mu \nu \alpha \beta}\left\langle \lambda L_\mu \right\rangle
\left\langle L_\nu L_\alpha L_\beta \right\rangle \right. \no \\*
& & +~ \left\langle \lambda [U^\dg \tilde{F}^{\mu\nu}_R U, L_\mu L_\nu]
\right\rangle \no \\*
& & +~ 3 \left\langle \lambda L_\mu \right\rangle \left\langle
(\tilde{F}^{\mu\nu}_L
+ U^\dg \tilde{F}^{\mu\nu}_R U) L_\nu \right\rangle \no \\*
& & + \left. \left\langle \lambda L_\mu \right\rangle \left\langle
(\tilde{F}^{\mu\nu}_L
- U^\dg \tilde{F}^{\mu\nu}_R U) L_\nu \right\rangle \right) .
\label{Qfac} \eeqa

Comparison with the general weak Lagrangian $\cL_4^{\Delta S=1}$ of
$O(p^4)$ in (\ref{L4w}), (\ref{Wi}) shows that all the possible octet operators
proportional to the $\ve$ tensor ($W_{28}$, $W_{29}$, $W_{30}$ and
$W_{31}$) appear in $Q_-$(fact) in (\ref{Qfac}). Thus, in a slightly
counter-intuitive way, the chiral anomaly contributes to all the
coefficients $N_{28},\ldots,N_{31}$ of normal octet operators.
Moreover,
the non-factorizable parts, which automatically have the
right octet transformation property
(they do not get any contribution from the anomaly),
must be of the same form
(\ref{Qfac}). The corresponding coefficients will differ from those in
Eq. (\ref{Qfac}). In fact, they must depend on the QCD scale $\mu$ to
cancel the $\mu$-dependence of the Wilson coefficients in the
$\Delta S =1$ effective Hamiltonian \cite{PR}.

Since all octet operators in $\Ha_{\rm eff}^{\Delta S=1}$ produce the
same structure (\ref{Qfac}), the $\Delta S=1$ effective Lagrangian in
the anomalous parity sector of $O(p^4)$ can be characterized by the
coefficients \cite{BEP1}
\beq
\ba{ll}
N_{28}^{\rm an} = \dfrac{a_1}{8\pi^2} \qquad \qquad &
N_{29}^{\rm an} = \dfrac{a_2}{32\pi^2} \\[10pt]
N_{30}^{\rm an} = \dfrac{3a_3}{16\pi^2} \qquad \qquad &
N_{31}^{\rm an} = \dfrac{a_4}{16\pi^2} .
\ea  \label{Nan}
\eeq
{}From the dominance of the octet operator $Q_-$, we expect the dimensionless
coefficients $a_i$ to be positive and of order 1. Unlike in the normal
parity sector at $O(p^2)$ \cite{PR}, the dominant penguin operator $Q_6$
does not contribute to the coefficients (\ref{Nan}) in the factorizable
approximation because there are no (pseudo-) scalar external fields in
$S_{WZW}$. Since $Q_6$ contributes constructively to the $O(p^2)$ weak
coupling $G_8$ \cite{PR}, which is pulled out in the definition of
$\cL_4^{\Delta S=1}$ in (\ref{L4w}), we expect the $a_i$ to be actually
smaller than one. The enhancement at $O(p^4)$ of the $\Delta I = 1/2$
$K \ra 2\pi$ amplitudes \cite{KMW2} lends additional support to this
expectation.

We are now in a position to determine all couplings relevant to
non-leptonic $K$ decays to which the chiral anomaly contributes in a
direct way via $\cL_4^{\Delta S=1}$ in (\ref{L4w}), (\ref{Wi})
with coefficients (\ref{Nan}). Restricting our attention to kinematically
allowed $K$ decays ($\leq 3$ pions, any number of photons), we obtain
\beqa
W_{28} &=& \frac{12\sqrt{2}\,i}{F^4} \ve_{\mu\nu\rho\sigma} \partial^\mu
K^0 \partial^\nu \pi^0 D^\rho \pi^+ D^\sigma \pi^- + \ldots \no \\
W_{29} &=& \frac{4ie}{F^3} \wt F_{\mu\nu} \left\{ 3K^+
\partial^\mu \pi^0 D^\nu
\pi^- + \sqrt{2} \, K^0 \left( D^\mu \pi^+ D^\nu \pi^- +
\frac{ie}{2} F^{\mu\nu} \pi^+ \pi^- \right)\right\} \no \\
&& \mbox{} + \frac{e}{F^4} \wt F_{\mu\nu} \{6K^+ \pi^0 D^\mu \pi^-
\partial^\nu \pi^0 - 4D^\mu K^+ D^\nu \pi^- \pi^+ \pi^- +
4K^+ \pi^- D^\mu \pi^+ D^\nu \pi^-
\quad
\no \\
&& \mbox{} + 3\sqrt{2}\, \partial^\mu K^0 \partial^\nu \pi^0 \pi^+ \pi^-
+ \sqrt{2}(4 K^0 \partial^\mu \pi^0 - \partial^\mu K^0 \pi^0)
(\pi^-
\stackrel{\leftrightarrow}{D^\nu} \pi^+)\} + \ldots \no \\
W_{30} &=& \frac{4ie}{F^3} \wt F_{\mu\nu} K^+ D^\mu \pi^- \partial^\nu
\pi^0 \no \\
&& \mbox{} + \frac{e}{F^4} \wt F_{\mu\nu} \{-2K^+ \pi^0 D^\mu \pi^-
\partial^\nu \pi^0  - 5\sqrt{2}\, \partial^\mu K^0 \partial^\nu
\pi^0 \pi^+ \pi^- \no \\
&& \mbox{} + \sqrt{2}\,  K^0 \partial^\mu \pi^0 (\pi^+
\stackrel{\leftrightarrow}{D^\nu} \pi^-) \} + \ldots \no \\
W_{31} &=& \frac{2\sqrt{2}\,ie}{F^3} \wt F_{\mu\nu} \partial^\mu K^0 (\pi^-
\stackrel{\leftrightarrow}{D^\nu} \pi^+) \no  \\
&& \mbox{} + \frac{e}{F^4} \wt F_{\mu\nu} \{ 2(K^+
\stackrel{\leftrightarrow}{D^\mu}
\pi^-) (\pi^+ \stackrel{\leftrightarrow}{D^\nu} \pi^-)  + \sqrt{2}\,
(\pi^0 \stackrel{\leftrightarrow}{\partial^\mu} K^0) (\pi^+
\stackrel{\leftrightarrow}{D^\nu} \pi^-)\}  + \ldots
\quad
\label{WiK}
\eeqa
We collect this information in Table \ref{taban} where all local
contributions from either the Lagrangian $\cL_{\rm an}^{\Delta S=1}$ in
(\ref{Law}) or the direct terms of $O(p^4)$ to all kinematically
allowed non-leptonic $K$ decays are listed. A separate column indicates
whether the corresponding decay has been observed experimentally. We
emphasize that the transitions with either three pions and/or two
photons in the final state are in general also subject to non-local
reducible anomalous contributions of the type shown in Fig. \ref{redaa}.

Finally, we observe that in the non-leptonic weak sector
the chiral anomaly contributes only to radiative $K$ decays.
\begin{table}
\caption{A complete list of local anomalous non-leptonic weak $K$ decay
amplitudes of $O(p^4)$ in the limit of CP conservation.} \label{taban}
$$
\begin{tabular}{|l|cccccc|} \hline
Transition & $\cL_{\rm an}^{\Delta S=1}$ & $W_{28}$ & $W_{29}$ & $W_{30}$
& $W_{31}$ & expt. \\ \hline
$K^+ \ra \pi^+ \pi^0 \gamma$       & x &   & x & x &   & x \\
$K^+ \ra \pi^+ \pi^0 \gamma\gamma$ & x &   & x & x &   &    \\
$K_L \ra \pi^+ \pi^- \gamma$       &   &   & x &   & x & x  \\
$K_L \ra \pi^+ \pi^- \gamma\gamma$ & x &   & x &   & x &    \\
$K^+ \ra \pi^+ \pi^0 \pi^0\gamma$  &   &   & x & x &   & x  \\
$K^+ \ra \pi^+ \pi^0 \pi^0 \gamma\gamma$ & & & x & x & & \\
$K^+ \ra \pi^+ \pi^+ \pi^- \gamma$ &   &   & x &   & x & x \\
$K^+ \ra \pi^+ \pi^+ \pi^- \gamma\gamma$ & & & x & & x & \\
$K_L \ra \pi^+ \pi^- \pi^0\gamma$ & & x & x & x && \\
$K_S \ra \pi^+ \pi^- \pi^0\gamma(\gamma)$ & & & x & x & x &  \\ \hline
\end{tabular}
$$
\end{table}

\renewcommand{\theequation}{\arabic{section}.\arabic{equation}}
\setcounter{equation}{0}
\setcounter{section}{3}
\section{General features of $K \ra \pi \pi \gamma$ decays}
The amplitude for $K(P) \ra \pi_1(p_1) + \pi_2(p_2) + \gamma(q)$
is decomposed into an electric amplitude $E(x_i)$ and a magnetic
amplitude $M(x_i)$:
\beq
A(K \ra \pi \pi\gamma) = \ve^\mu(q)^* [E(x_i)(p_1 q p_{2\mu} -
p_2 q p_{1\mu}) + M(x_i) \ve_{\mu\nu\rho\sigma} p_1^\nu p_2^\rho
q^\sigma]/M_K^3 \label{adec}
\eeq
\nopagebreak
$$
x_i = \frac{P p_i}{M_K^2} \qquad (i = 1,2), \qquad
x_3 = \frac{Pq}{M_K^2}, \qquad x_1 + x_2 + x_3 = 1.
$$
The invariant amplitudes $E(x_i)$, $M(x_i)$ are dimensionless. Summing
over the photon helicity, the differential decay distribution can be
written as ($r_i = M_{\pi_i}/M_K$)
\beqa
\frac{\partial^2\Gamma}{\partial x_1 \partial x_2} &=&
\frac{M_K}{4(4\pi)^3} (|E(x_i)|^2 + |M(x_i)|^2)
[(1 - 2x_3 - r_1^2 - r_2^2)(1 - 2x_1 + r_1^2 - r_2^2) \cdot \no \\
&& (1 - 2x_2 + r_2^2 - r_1^2) - r_1^2(1 - 2x_1 + r_1^2 - r_2^2)^2
- r_2^2(1 - 2x_2 + r_2^2 - r_1^2)^2] .\no \\
&&
\eeqa
There is no interference between $E$ and $M$ as long as the photon helicity
is not measured. In the following, we will not include the strong
$\pi \pi$ rescattering phases in the amplitudes $E$, $M$ \cite{CK}. Of
course, those phases should and are usually taken into account in the
experimental analysis.

For most of the $K \ra \pi \pi \gamma$ decays, the electric amplitude is
dominated by the bremsstrahlung amplitude $E_B(x_i)$. This amplitude
arises already at lowest $O(p^2)$ in CHPT. In fact, the following theorem
\cite{ENP1,RafRing} shows that to $O(p^2)$ the $K \ra \pi\pi\gamma$
amplitudes (actually $K \ra \pi \pi (n\gamma)$ for any $n \geq 1$) are
completely determined by $E_B(x_i)$. In other words, there is no additional
information to $O(p^2)$ that would not already be contained in the
corresponding non-radiative transitions $K \ra \pi\pi$.

\paragraph{Theorem:} Consider a general Lagrangian $\cL_2(\vp_i,D_\mu \vp_i)$
$(i = 0,+,-)$ with at most two (electromagnetically gauge covariant)
derivatives. In addition to the kinetic terms, there are only cubic
interactions. Then the tree level amplitude for $\vp_0,\vp_+,\vp_-$
and any number $n$ of photons in the initial or final states factorizes,
\beq
A(\vp_0\vp_+\vp_- \gamma_1 \ldots \gamma_n) =
A_B(\ve_a,q_a,p_i) A(\vp_0 \vp_+ \vp_-) \qquad a = 1,\ldots,n \qquad
i = 0,+,-. \label{BTh}
\eeq
$A(\vp_0\vp_+\vp_-)$ is the on-shell amplitude for the decay of either
spin-0 particle into the other two and $A_B(\ve_a,q_a,p_i)$ is the
general bremsstrahlung amplitude independent of the structure of
$\cL_2$. \\

The proof is straightforward and is relegated to Appendix A. Here, we add a
few clarifying comments.
\begin{enumerate}
\item[i.] Although the notation is suggestive, the conclusion is not
restricted to $O(p^2)$ in CHPT. Arbitrary mass terms in the chiral
expansion fall under the general assumptions as long as there are at
most two derivatives in the respective couplings. As a particular
consequence, parts of the $O(p^4)$ CHPT corrections are covered by the
theorem.
\item[ii.] The same constraints of gauge invariance and at most two
derivatives imply that the amplitude $K \ra \pi \gamma^* \ldots \gamma^*$
vanishes for any number of real or virtual photons \cite{EPR3}. There,
only the gauge-invariant kinetic parts enter.
\item[iii.] A corresponding statement does not hold for more than three
particles or more than two derivatives. We shall come back to this remark
in Sect.~6. Note that this comment does not contradict Low's theorem
\cite{Low}, which is of course always valid.
\item[iv.] Although relatively trivial for $n = 1$, the relation
(\ref{BTh}) can save a considerable amount of work for $n \geq 2$.
\end{enumerate}

In the next section, we will try to estimate the dominant effects of
$O(p^6)$ for the transitions $K_L \ra \pi^+ \pi^- \gamma$ and
$K^+ \ra \pi^+ \pi^0 \gamma$ due to vector meson exchange. As is the
case in general for non-leptonic weak transitions, there are two
different mechanisms related to V exchange \cite{EPR4}. The first
mechanism involves a strong VMD amplitude in connection with a
non-leptonic weak transition on the external pseudoscalar meson
legs. Given the strong amplitude, the weak VMD amplitude is unambiguously
calculable through a weak rotation \cite{EPR3}. This is unfortunately not
the case for the so-called direct weak terms corresponding
to the weak Lagrangian $\cL_6^{\Delta S=1}$ in the present situation.
Even at $O(p^4)$, one must resort to models to obtain estimates of such
terms related to V exchange \cite{EKW1}.

In Ref. \cite{ENP1}, the so-called weak deformation model (WDM)
\cite{EPR4} was used to estimate the direct weak terms for the magnetic
amplitudes of $O(p^6)$. Another model that has been used frequently in
non-leptonic weak transitions (see \cite{PR} and references quoted
therein) is the factorization model (FM). The FM is motivated by
large-$N_C$ arguments\footnote{A more systematic treatment of the
large-$N_C$ expansion in this connection can be found in Ref.
\cite{BP}.} and can be defined as (keeping only the octet part)
\beq
\cL_{\rm FM} = 4 k_f G_8 \left\langle \lambda \frac{\delta S}
{\delta \ell_\mu} \frac{\delta S}{\delta \ell^\mu} \right\rangle
+ {\rm h.c.} \label{LFM}
\eeq
where $S$ is the CHPT action for the strong interactions and
\beq
\frac{\delta S}{\delta \ell_\mu} =: J_L^\mu = J^\mu_{L,1} +
J^\mu_{L,3} + J^\mu_{L,5} + \ldots \qquad \qquad
J^\mu_{L,1} = - \frac{i}{2} F^2 U^\dg D^\mu U
\eeq
is the corresponding left-chiral current. The constant $k_f$ is a fudge
factor which the na\"{\i}ve FM puts equal to one.\footnote{For the weak
anomalous action of Sect. 3, $k_f = 1$ corresponds to $a_i =1$
$(i = 1,\ldots,4)$ in Eq. (\ref{Nan}).} As shown in Ref. \cite{EKW1},
the WDM can be expressed through the Lagrangian
\beq
\cL_{\rm WDM} = 2 G_8 \left\langle \lambda \left\{ J^\mu_{L,1},
\frac{\delta S}{\delta \ell^\mu}\right\} \right\rangle + {\rm h.c.}
\label{LWDM}
\eeq
This Lagrangian formulation of the WDM immediately leads to the result
\cite{EKW1} that $\cL_{\rm WDM}$ is a special case of $\cL_{\rm FM}$
for $k_f = 1/2$ to $O(p^4)$. However, starting at $O(p^6)$ the FM has
additional terms not contained in the WDM.

For the transitions of interest here, we are only concerned with the
magnetic amplitudes of $O(p^6)$. Since the strong action $S$ of
$O(p^4)$ \cite{GL2} has no terms with an $\ve$ tensor except for the
anomaly, the relevant FM Lagrangian consists of the following two
terms:
\beq
\cL_{\rm FM}(M_6) = 4 k_f G_8 \left\{
\left\langle \lambda \left\{ \frac{\delta S_2}{\delta \ell_\mu},
\frac{\delta S_6}{\delta \ell^\mu}\right\} \right\rangle +
\left\langle \lambda \left\{ \frac{\delta S_4}{\delta \ell_\mu},
\frac{\delta S_{\rm WZW}}{\delta \ell^\mu}\right\} \right\rangle
\right\} + {\rm h.c.}
\label{FM6}
\eeq
Comparing with $\cL_{\rm WDM}$ in (\ref{LWDM}), the first term of
$\cL_{\rm FM}(M_6)$ reduces again to the WDM for $k_f = 1/2$ since
\beq
\frac{\delta S_2}{\delta \ell_\mu} = J^\mu_{L,1}.
\eeq
We can therefore simply multiply the WDM amplitudes derived in Ref.
\cite{ENP1} by $2k_f$ to get the corresponding FM amplitudes.

The second term in Eq. (\ref{FM6}) involves the anomalous current
(\ref{Jan}) and the normal current $\delta S_4/\delta \ell_\mu$ of
$O(p^3)$. It is well-known \cite{EGPR} that the dominant terms in $S_4$
are due to spin-1 exchange. Taking the special form of the anomalous
current in Eq. (\ref{Jan}) into account and restricting ourselves to
couplings sensitive to spin-1 exchange \cite{EGPR}, one finds that a
single term in $S_4$, albeit the one with the biggest coupling constant
$L_9$, can contribute to $K \ra \pi \pi \gamma$ decays via
(\ref{FM6}). Omitting all terms irrelevant for our transitions,
the (matrix) current of $O(p^3)$ is given by
\beq
\frac{\delta S_4}{\delta \ell^\mu} = i L_9 \partial^\nu
(\partial_\nu U^\dg \partial_\mu U - \partial_\mu U^\dg \partial_\nu U)
+ \ldots \label{J3n}
\eeq

\renewcommand{\theequation}{\arabic{section}.\arabic{equation}}
\setcounter{equation}{0}
\setcounter{section}{4}
\section{Phenomenology of $K \ra \pi\pi \gamma$ decays}
{}From the analysis of Sect. 3 summarized in Table \ref{taban}, the
chiral anomaly is seen to contribute only to the decays
$K^+ \ra \pi^+ \pi^0 \gamma$ and $K_L \ra \pi^+ \pi^- \gamma$ at
$O(p^4)$. In this section, we perform a detailed phenomenological
analysis of these decays. For completeness, we include some remarks
about the remaining $K \ra \pi \pi \gamma$ decays referring to and
commenting on recent work.

\subsection{$K_{L,S} \ra \pi^0 \pi^0 \gamma$}
For the decays $K^0 \ra \pi^0 \pi^0 \gamma$, Bose statistics implies
\beqa
E(x_2,x_1) &=& - E(x_1,x_2) \no \\
M(x_2,x_1) &=& - M(x_1,x_2). \label{Bose}
\eeqa
In the limit where CP is conserved, the amplitude for $K_L$ ($K_S$) is
purely electric (magnetic).

The transition $K_L \ra \pi^0 \pi^0 \gamma$ has recently been considered
in the literature \cite{FK,HS2}.
Eq.~(\ref{Bose}) implies the absence
of a local amplitude of $O(p^4)$, or more generally the absence of an E1
amplitude. Although this by itself does not imply a vanishing one-loop
amplitude (as can be seen in the case of $K_L \ra \pi^+ \pi^- \gamma$
later in this section), Funck and Kambor \cite{FK} have shown that it
does indeed vanish for a real photon. For a virtual photon, the one-loop
amplitude is non-zero. In fact, it is divergent and it gets renormalized
by the same combination of weak counterterms \cite{FK}
\beq
2N_{14} + N_{15}
\eeq
appearing in the transition $K_1^0 \ra \pi^0 \gamma^*$ \cite{EPR1}.

Thus, the decay $K_L \ra \pi^0 \pi^0 \gamma$ with a real photon is at
least $O(p^6)$ in CHPT. In fact, chiral symmetry permits local octet
couplings of $O(p^6)$ contributing to this transition. A typical term,
compatible with all symmetries, is provided by \footnote{For our purposes,
the covariant derivatives in (\ref{K6}), (\ref{LK6}) can be replaced
by normal ones.}
\beq
\frac{1}{2} \left\langle \{ \Delta,f_+^{\mu\nu}\} (\nabla_\lambda u_\mu
u^\lambda u_\nu + u_\nu u^\lambda \nabla_\lambda u_\mu)\right\rangle.
\label{K6}
\eeq
A survey of vector meson couplings of $O(p^3)$ \cite{EGLPR} shows that
there is no strong amplitude of $O(p^6)$ induced by V exchange that
could contribute to $K_L \ra \pi^0 \pi^0 \gamma$ via a weak rotation.
It therefore seems legitimate to estimate the strength of the coupling
(\ref{K6}) by na\"{\i}ve chiral dimensional analysis \cite{GM} as
\beq
\cL_6^{\Delta S=1} = \frac{G_8}{2(4\pi)^4}
\left\langle \{ \Delta,f_+^{\mu\nu}\} (\nabla_\lambda u_\mu
u^\lambda u_\nu + u_\nu u^\lambda \nabla_\lambda u_\mu)\right\rangle
+ {\rm h.c.} +\ldots \label{LK6} \eeq

The corresponding amplitude for $K_L \ra \pi^0 \pi^0 \gamma$ is
\beq
E_6(x_1,x_2) = \frac{4 i G_8 e M_K^5}{3(4\pi)^4 F^3}(x_1 - x_2) \, ,
\eeq
yielding a branching ratio
\beq
\left. BR(K_L \ra \pi^0 \pi^0 \gamma) \right|_{O(p^6)} = 7 \cdot 10^{-11}.
\eeq
By relating $K_L \ra \pi^0 \pi^0 \gamma$ to the decay
$K_L \ra \pi^+ \pi^- \gamma$ (which is dominantly M1), Heiliger and Sehgal
obtain a considerably bigger estimate \cite{HS2}
$\left. BR(K_L \ra \pi^0 \pi^0 \gamma)\right|_{\rm HS} = 1 \cdot 10^{-8}$
together with
$\left. BR(K_S \ra \pi^0 \pi^0 \gamma)\right|_{\rm HS} = 1.7 \cdot
10^{-11}.$

\subsection{$K_S \ra \pi^+ \pi^- \gamma$}
In the limit of CP conservation, the amplitudes for $K_S \ra \pi^+ \pi^-
\gamma$ obey the symmetry relations
\beqa
E(x_-,x_+) &=& E(x_+,x_-)  \label{KSCP} \\
M(x_-,x_+) &=& -M(x_+,x_-). \no
\eeqa
To $O(p^4)$, the amplitude is therefore purely electric.  In addition to
the bremsstrahlung amplitude of $O(p^2)$ (cf. theorem (\ref{BTh})), the
loop and counterterm amplitudes of $O(p^4)$ have recently been calculated
by D'Ambrosio, Miragliuolo and Sannino \cite{DAMS}. The local
contribution of $O(p^4)$ is proportional to
\beq
N_{14} - N_{15} - N_{16} - N_{17}
\eeq
and it is scale-independent \cite{KMW1,EKW1}.
The same combination of coupling constants appears in the electric
amplitude for the decay $K^+ \ra \pi^+ \pi^0 \gamma$ \cite{ENP1}.
Consequently, the loop amplitudes for both $K_S \ra \pi^+ \pi^- \gamma$
and $K^+ \ra \pi^+ \pi^0 \gamma$ are finite.

At present, experimental data \cite{RPP} are consistent with a pure
bremsstrahlung amplitude. However, forthcoming facilities like DAPHNE
\cite{DAF} should be able to detect interference with the $O(p^4)$
amplitude that is expected to show up at the level of $10^{-6}$ in
branching ratio (for $E_\gamma > 20$~MeV) \cite{DAMS}.

\subsection{$K_L \ra \pi^+ \pi^- \gamma$}
The bremsstrahlung amplitude of $O(p^2)$ \cite{ENP1}
\beq
E_B(x_i) = \frac{\ve e A(K_1^0 \ra \pi^+\pi^-)}{M_K(\frac{1}{2}-x_+)
(\frac{1}{2}-x_-)} \qquad \qquad p_1 = p_+, \quad p_2 = p_- \label{EBREMS}
\eeq
violates CP. Here $\ve$ is the standard CP violation parameter in
$K \ra \pi\pi$ decays and we have neglected $\ve'$. From
$O(p^4)$ on we assume CP conservation implying [cf. Eq. (\ref{KSCP})]
\beqa
E(x_-,x_+) &=& -E(x_+,x_-) \no \\
M(x_-,x_+) &=& M(x_+,x_-). \label{KLCP}
\eeqa
The dominant contribution of $O(p^4)$ occurs in the magnetic amplitude
and it is due to the anomaly. As discussed in Sect. 3, there is no
reducible anomalous amplitude of $O(p^4)$. The direct weak
anomaly functional gives rise to \cite{BEP1,Cheng}
\beq
M_4 = \frac{e G_8 M_K^3}{2\pi^2 F} (a_2 + 2a_4) \label{KLM4}
\eeq
in terms of the coupling constants $a_i$ defined in Eq. (\ref{Nan}).

Because of (\ref{KLCP}) there is no local contribution to $E$ at
$O(p^4)$. In contrast to $K_L \ra \pi^0 \pi^0 \gamma$, there is
however a finite one-loop amplitude. The relevant Feynman diagrams are
shown in Fig. \ref{KLfig}.
The result of the loop calculation is proportional to the non-leptonic
weak vertex occurring in Fig. \ref{KLfig}, where the momenta of the
corresponding three mesons are put on the mass shell. Consequently, only
the diagrams of type $b$ give non-vanishing amplitudes for the
$\pi^\pm K_1^0$ and the $K^\pm \eta$ intermediate states.
In accordance with (\ref{KLCP}), the loop amplitude for
$K_L \ra \pi^+ \pi^- \gamma$ takes the form \cite{ENP1}
\beq
E_4^{\rm loop}(x_+,x_-) = \frac{i e G_8 M_K(M_K^2 - M_\pi^2)}{8\pi^2 F}
[g(x_-) - g(x_+)] \, , \label{gx}
\eeq
where the function $g(x)$ is defined in Appendix B.
This result can now be
compared with the bremsstrahlung amplitude (\ref{EBREMS}):
\begin{figure}
\epsfig{file=f2anom.eps,height=7.0cm}
\caption{One-loop diagrams for $K_2^0 \ra \pi^+ \pi^- \gamma$. The
photon is to be appended on all charged lines and on both the
non-leptonic weak vertex (square) and the strong vertex (circle).
Tadpole diagrams do not contribute to the amplitude.}
\label{KLfig}
\end{figure}
\beq
\left|\frac{E_4^{\rm loop}}{E_B}\right| = \left| \frac{M_K^2}
{\ve(4\pi F)^2} [g(x_-) - g(x_+)] \left(\frac{1}{2} - x_+\right)
\left(\frac{1}{2} - x_-\right)\right|.
\eeq
Taking the maximum of this ratio over the whole Dalitz plot leads
to the bounds
\beqa
\left| \frac{E_4^{\rm loop}(\pi^\pm K_1^0)}{E_B} \right| &\leq&
1.1 \cdot 10^{-2} , \no \\
\left| \frac{E_4^{\rm loop}(K^\pm \eta)}{E_B} \right| &\leq&
0.1 \cdot 10^{-2}
\eeqa
for the contributions of the $\pi^\pm K_1^0$ and $K^\pm \eta$ intermediate
states, respectively. The smallness of the ratio of the two amplitudes
is, of course, due to CP invariance, which is responsible for the
antisymmetry in $x_+$, $x_-$ of $E_4^{\rm loop}$, forbidding in
particular an electric dipole amplitude. Note also that because of
$\arg \ve \simeq \pi/4$ there is only partial interference between
$E_4^{\rm loop}$ and $E_B$. It seems almost impossible to
detect the loop amplitude.

For the electric amplitude $E$, the analysis to $O(p^4)$ is therefore
more than sufficient. There are on the other hand strong experimental
indications for the presence of a sizeable magnetic amplitude beyond
$O(p^4)$. A recent analysis of $K_L \ra \pi^+ \pi^- \gamma$ at FNAL
\cite{E731} confirms an earlier result from Brookhaven \cite{BNL}
finding evidence for a dependence of the direct emission amplitude on
the photon energy. On the other hand, the dominant direct emission
amplitude $M_4$ in (\ref{KLM4}) is a constant, independent of the photon
energy.

At $O(p^6)$, CP invariance leads to the following most general form
of the magnetic amplitude via Eq. (\ref{KLCP}):
\beq
M_6(x_+,x_-) = a + b(x_+ + x_-) = a + b - b x_3 = \wh M_6 - b x_3,
\qquad x_3 = \frac{Pq}{M^2_K} = \frac{E_\gamma}{M_K},
\label{x3}
\eeq
where $E_\gamma$ is the photon energy in the kaon rest frame. To
$O(p^6)$, the total magnetic amplitude is therefore given by
\beq
M(x_+,x_-) = M_4 + M_6(x_+,x_-) = M_4 + \wh M_6 - b x_3 =
(M_4 + \wh M_6)(1 + c x_3) \label{defc}.
\eeq
{}From the distribution in $E_\gamma$ measured by E731 \cite{E731}, one
can extract \cite{Ram} a value
\beq
c = - 1.7 \pm 0.5 \label{cRam}
\eeq
for the slope $c$, in agreement with the earlier measurement
\cite{BNL,LV}.

How can CHPT account for this rather big slope? An early explanation
was put forward by Lin and Valencia \cite{LinV}, who suggested a
vector-meson-dominated
form factor in the $\pi^+ \pi^-$ invariant mass to be
responsible for the slope. The experimental value (\ref{cRam}) of the
slope is in fact consistent with their model amplitude. Unfortunately,
as already noted by Picciotto \cite{Picc}, the amplitude of Ref.
\cite{LinV} violates chiral symmetry. In the terminology of Sect. 3,
their magnetic amplitude is of the reducible type, corresponding in
particular to $a_i = 0$ $(i = 1,\ldots,4)$. To agree with our general
result (\ref{KLM4}), their amplitude should therefore vanish at
$O(p^4)$, which in fact it does not. The source of the problem seems
to lie \cite{Picc} in the model for combining the chiral anomaly and
vector mesons.

Vector meson exchange contributes first at $O(p^6)$ to the amplitude
$M$. This implies that the dependence on $E_\gamma$ due to a $V$
propagator is an effect of $O(p^8)$ and higher. Although not
impossible, the big slope of Eq. (\ref{cRam}) makes the interpretation
as an $O(p^8)$ effect difficult to understand for the chiral
practitioner. To make this feeling more quantitative, let us adopt
the simplifying assumption that $M_6$ in (\ref{x3}) is entirely due to
$V$ exchange. In this case
\beq
M_6(x_+,x_-) = \frac{M_6^0}{1 - \dfrac{(p_++p_-)^2}{M_V^2}} =
M_6^0 \left[ 1 + \frac{M_K^2}{M_V^2} (1-2x_3) + \ldots\right]
\quad (M_V \simeq M_\rho) \label{M6V}
\eeq
and consequently
\beq
\wh M_6 = M_6^0 \left( 1 + \frac{M_K^2}{M_V^2}\right), \qquad
c = - \frac{2 M_K^2 M_6^0}{M_V^2(M_4 + \wh M_6)}.
\eeq
Making the plausible assumption that the amplitude $\wh M_6$ is at most
equal to $M_4$ in magnitude, the observed sign of $c$ requires that
$M_4$ and $\wh M_6$ (or $M_6^0$) interfere constructively. Moreover,
with $|M_4| > |\wh M_6|$ the absolute value of the slope is bounded by
\beq
|c| = \frac{2M_K^2 M_6^0/\wh M_6}{M_V^2(1 + M_4/\wh M_6)} <
\frac{M_K^2}{M_V^2 + M_K^2} = 0.3,
\eeq
much too small to explain the measured value (\ref{cRam}).

We are therefore led to interpret the slope $c$ as an effect of $O(p^6)$.
Which are the dominant contributions of $O(p^6)$? First of all, there is
a reducible amplitude due to the anomaly of the form
\beq\label{fterm}
M_6^{\rm anom} = - \frac{e G_8 M_K^3}{2\pi^2 F} F_1 \label{M6an}
\eeq
$$
F_1 = \frac{1}{1 - r_\pi^2} -
\frac{(c - \sqrt{2}\,s)(c + 2\sqrt{2}\,\rho s)}{3(r_\eta^2 - 1)} +
\frac{(\sqrt{2}\,c + s)(2\sqrt{2}\,\rho c -s)}{3(r_{\eta'}^2 -1)}
$$
$$
r_i = M_i/M_K, \qquad c = \cos \Theta, \qquad s = \sin \Theta
$$
in the notation of Ref. \cite{ENP1}; $\Theta$ denotes the $\eta$--$\eta'$
mixing angle and $\rho \neq 1$ takes into account possible deviations
from nonet symmetry for the non-leptonic weak vertices (nonet symmetry
is assumed for the strong WZW vertices).
At $O(p^4)$ ($\Theta = 0$, $M_{\eta'} \ra \infty$), $F_1$ vanishes
because of the Gell-Mann--Okubo mass formula. In the real world, the
$\eta$ and $\eta'$ contributions interfere destructively for
$0 \leq \rho < 1$ and $\Theta \simeq - 20^\circ$ as in the similar case of
the $K_L \ra 2\gamma$ amplitude. Although not really predictable with any
precision, $F_1$ is dominated by the pion pole and certainly positive.
We observe that $M_4$ (with $a_i > 0$) and $M_6^{\rm anom}$ interfere
destructively, as already noted by Cheng \cite{Cheng}, making a
reliable estimate all the more difficult. At the present state of the art,
the real challenge in $K_L \ra \pi^+ \pi^- \gamma$ is to understand the
sign and magnitude of the slope $c$.

In the simplifying limit $M_\pi = 0$, the strong VMD amplitude of
$O(p^6)$ is unique. The relevant couplings are defined by the Lagrangian
\cite{EGLPR}
\beq
\cL_V = - \frac{ig_V}{2\sqrt{2}} \langle \wh V_{\mu\nu}[u^\mu,u^\nu]\rangle
+ h_V \langle \wh V_\mu \{ u_\nu,\wt f_+^{\mu\nu}\}\rangle + \ldots
\qquad \wh V_{\mu\nu} = \nabla_\mu \wh V_\nu - \nabla_\nu \wh V_\mu
\eeq
for the vector meson resonance field $\wh V_\mu$. Contracting the vector
meson fields to produce an effective strong VMD Lagrangian of $O(p^6)$
proportional to $g_V h_V$ and applying a weak rotation leads to the weak
VMD amplitude \cite{ENP1} ($M_\pi = 0$)
\beq
M_6^{\rm VMD} = 2 C_V (1 - 3x_3), \qquad \qquad
C_V = \frac{16 \sqrt{2} \, e G_8 g_V h_V M_K^5}{3 M_V^2 F}.
\eeq

To estimate the direct weak amplitude of $O(p^6)$ related to $V$ exchange,
we make use of the FM as discussed in the previous section. With the
Lagrangian defined in Eq. (\ref{FM6}), one obtains ($M_\pi = 0$)
\beq
M_6^{\rm FM} = 4 k_f C_V x_3 + \frac{e G_8 M_K^5 L_9}{\pi^2 F^3} k_f
(2 - 5x_3).
\eeq
The first term reduces to the WDM amplitude for $k_f = 1/2$ \cite{ENP1},
whereas the second term has no analogue in the WDM and is proportional to
the $O(p^4)$ coupling constant $L_9$ appearing in the current
(\ref{J3n}).

Altogether, we obtain for the magnetic amplitude
\beq
M(x_3) = \frac{eG_8 M_K^3}{2\pi^2 F} \left\{ a_2 + 2a_4 - F_1 +
r_V [1 + x_3(2k_f - 3)] + \frac{2L_9 M_K^2}{F^2} k_f(2 - 5x_3)\right\}
\label{MKL}
\eeq
$$
r_V = \frac{64 \sqrt{2} \,\pi^2 g_V h_V M_K^2}{3 M_V^2} \simeq 0.4
\simeq \frac{2 L_9^r(M_\rho) M_K^2}{F_\pi^2}
$$
with \cite{EPR4,EGLPR,GL2}
\beq
g_V \simeq \frac{F_\pi}{\sqrt{2}\,M_V}, \qquad
|h_V| \simeq 3.7 \cdot 10^{-2}, \qquad
L_9^r(M_\rho) \simeq 6.9 \cdot 10^{-3}.
\eeq
In contrast with more phenomenologically oriented treatments, CHPT as a
quantum field theory permits a reliable determination of the relative
signs of the various terms in the amplitude (\ref{MKL}):
\begin{itemize}
\item Although we cannot predict a precise value for the quantity
$a_2 + 2a_4 - F_1$, factorization discussed in Sect.~3
($0 < a_i \; \letsmod \; 1$) strongly indicates a positive sign.
\item Although the rates $\Gamma(V \ra P \gamma)$ only determine
$|h_V|$, the product $g_Vh_V$, and therefore $r_V$, must be positive.
The argument invokes yet another vector meson coupling constant $f_V$
\cite{EGPR,EGLPR}. The product $f_V h_V$ governs the slope of the
$\pi^0,\eta \ra \gamma\gamma^* \ra \gamma \ell^+ \ell^-$ amplitudes
in the virtual photon mass \cite{EGosen}. Experimental evidence (see
the discussion in Refs.~\cite{EGosen,BBC}) agrees with the
predicted magnitude and fixes $f_V h_V > 0$. On the other hand,
$f_V g_V \simeq F_\pi^2/M_V^2$ \cite{EGLPR} is known to be positive and
so is therefore $g_V h_V$ (see also Ref.\cite{Prades}).
\item $L_9^r(M_\rho)$ is certainly positive \cite{GL2}. In resonance
approximation \cite{EGPR,EGLPR} $L_9 = \frac{1}{2} f_V g_V$,
substantiating the previous argument.
\end{itemize}
Comparing the total magnetic amplitude (\ref{MKL}) with the definition
(\ref{defc}) of the slope parameter $c$, we infer that $c$ must be
negative for all reasonable values of the factorization parameter
$k_f$ ($0 < k_f \; \letsmod \; 1$). To find out whether (\ref{MKL}) can
also explain the magnitude of the experimentally measured slope
(\ref{cRam}), we use the recent measurement \cite{E731}
\beq
BR(E_\gamma > 20\mbox{ MeV})_{\rm DE} = (3.19 \pm 0.16)\cdot 10^{-5}
\label{BRKL}
\eeq
of the direct emission branching ratio to determine the quantity
$a_2 + 2a_4 - F_1$ for given values of $k_f$. Then, the slope $c$ can
be extracted from Eq. (\ref{MKL}) both in magnitude and sign.

The results are displayed in Table \ref{ctab} for three representative
values of $k_f$. The fitted values of $a_2 + 2a_4 - F_1$ document the
expected strong destructive interference between the leading term
$a_2 + 2a_4$ and the $O(p^6)$ correction $F_1$. Our main results are
the big values for $|c|$ as found experimentally. In view of
Eq. (\ref{M6V}), we may in addition expect an
enhancement of $|c|$ by the propagator effect of $O(p^8)$. However,
our analysis reinforces the previous conclusion that the slope
parameter is dominantly an effect of $O(p^6)$. In summary, we cannot
claim to be able to predict the rate for $K_L \ra \pi^+ \pi^- \gamma$,
but CHPT establishes a correlation between the rate and the slope
parameter $c$, in agreement with experiment.
\begin{table}
\caption{The slope parameter $c$ as a function of the factorization
parameter $k_f$. The quantity $a_2 + 2a_4 - F_1$ is extracted from
the measured branching ratio (\protect{\ref{BRKL}}).}
\label{ctab}
$$
\begin{tabular}{|c|c|c|} \hline
$k_f$ & $a_2 + 2a_4 - F_1$ & $c$ \\ \hline
0 & 0.9 & $- 0.9$ \\
0.5 & 0.6 & $- 1.3$ \\
1 & 0.3 & $- 1.6$ \\
\hline
\end{tabular}
$$
\end{table}

In addition to the analysis of Ref. \cite{LinV} already mentioned, several
authors have addressed the decay $K_L \ra \pi^+ \pi^- \gamma$ recently.
Cheng \cite{Cheng} used factorization for the $O(p^4)$ magnetic amplitude
($a_2 = a_4 = 1$). He has emphasized the need for a strong destructive
interference between the leading contribution and higher-order terms
like $F_1$, but he did not include $V$ exchange
($r_V = L_9 = 0$ in (\ref{MKL})).
In two more recent papers
\cite{KT,Picc}, vector meson exchange is included.
The magnetic amplitudes
of Ko and Truong \cite{KT} and of Picciotto \cite{Picc} agree to $O(p^6)$
with our amplitude (\ref{MKL}) in the (not very realistic) limit
$a_2 = a_4 = 0$ (no direct anomalous amplitude) and $k_f = 0$ (pure
VMD only). Within the hidden symmetry approach \cite{Bando} for the
``anomalous'' couplings of vector mesons, they go beyond $O(p^6)$ by
including in particular the vector meson propagators, but they find
essentially no dependence of $M$ on the photon energy.

Further work on $K \to \pi \pi \gamma$ decays can be found in
Ref. \cite{KPPG}.

\subsection{$K^+ \ra \pi^+ \pi^0 \gamma$}
The decay $K^+ \ra \pi^+ \pi^0 \gamma$ shares several features with
$K_L \ra \pi^+ \pi^-\gamma$:
\begin{itemize}
\item The bremsstrahlung amplitude is suppressed;
\item The dominating contribution of $O(p^4)$ is due to the chiral anomaly;
\item The one-loop amplitude is finite, but again very small.
\end{itemize}

The bremsstrahlung amplitude \cite{ENP1}
\beq
E_B(x_i) = \frac{e A(K^+ \ra \pi^+ \pi^0)}{M_K x_3(\frac{1}{2} -x_0)},
\qquad \qquad p_1 = p_+, \quad p_2 = p_0, \label{EBKP}
\eeq
includes the complete amplitude of $O(p^2)$ according to the theorem
of Sect.~4 and it is suppressed by the $\Delta I = 1/2$ rule.

The magnetic amplitude of $O(p^4)$ consists of both a reducible and a
direct amplitude \cite{ENP1,BEP1}:
\beq
M_4 = \frac{e G_8 M_K^3}{2 \pi^2 F} \left(-1 + \frac{3}{2} a_2 - 3a_3\right).
\label{M4Kp} \eeq
Factorization suggests constructive interference between these two
terms.

In contrast with $K_L \ra \pi^+ \pi^- \gamma$, there is now a local
scale-independent contribution of $O(p^4)$ to the electric amplitude
$E$ \cite{ENP1}:
\beq
E_4^{\rm local} = \frac{2ie G_8 M_K^3}{F} (N_{14} - N_{15} - N_{16}
- N_{17}). \label{E4KP}
\eeq
As already mentioned, the same combination of coupling constants
appears in the amplitude for $K_S \ra \pi^+ \pi^- \gamma$ \cite{DAMS}.
By measuring the energy spectrum of the photon, the counterterm
amplitude (\ref{E4KP}) can in principle be isolated through its
interference with the bremsstrahlung amplitude (\ref{EBKP}). We can
estimate the size\footnote{Although $N_{14} - N_{15}$ can be determined
from the recent measurement of $K^+ \ra \pi^+ e^+ e^-$ \cite{All,EPR1},
the constants $N_{16}$, $N_{17}$ are still unknown.}
of this interference by appealing to the FM which predicts \cite{EKW1}
\beq
N_{14} - N_{15} - N_{16} - N_{17} = - k_f \frac{F_\pi^2}{2 M_V^2} =
- 7 \cdot 10^{-3} k_f.
\eeq
For $k_f > 0$, the interference is predicted to be positive
\cite{ENP1}:
\beq
\frac{E_4^{\rm local}}{E_B} \simeq 2.3 x_3(1 - 2x_0)
(- N_{14} + N_{15} + N_{16} + N_{17})/7 \cdot 10^{-3}.
\eeq
The sign is well-determined because the ratio $G_8/G_{27}$ is known to be
positive from $K \ra 2\pi$ decays (see Sect.~2). Except for small
$E_\gamma$ ($x_3 \ra 0$, $2x_0 \ra 1$) where bremsstrahlung is bound to
dominate, the amplitude $E_4^{\rm local}$ should be detectable. In fact,
the experiment of Abrams et al. \cite{KPexp} is consistent with
constructive interference between $E_B$ and $E_4^{\rm local}$,
but the available data \cite{KPexp} are not precise enough
to separate the amplitudes $E - E_B$ and $M$ experimentally.

We now turn to the loop amplitude, which is necessarily finite. The
Feynman diagrams are shown in Fig. \ref{KPfig}. Similar to
$K_L \ra \pi^+ \pi^- \gamma$, only the graphs of type $b$ with
$\pi^+ K_1^0$ and $K^+ \eta$ intermediate states yield
non-vanishing contributions  in the octet limit ($G_{27} = 0$).
The corresponding amplitude is given by
\begin{figure}
\centerline{\epsfig{file=f3anom.eps,width=11.5cm}}
\caption{One-loop diagrams for $K^+  \ra \pi^+ \pi^0 \gamma$ (notation
as in Fig. \protect{\ref{KLfig}}).}
\label{KPfig}
\end{figure}
\beq
E_4^{\rm loop}(x_0) = \frac{ieG_8 M_K(M_K^2 - M_\pi^2)}{8\pi^2 F} h(x_0)
\label{H}
\eeq
where the function $h(x)$ can again be found in Appendix B.
The ratio of the loop
amplitude to the bremsstrahlung amplitude can now be written in the form
\beq
\left| \frac{E_4^{\rm loop}}{E_B}\right| = \left|
\frac{M_K^2 G_8}{24 \pi^2 F^2 G_{27}^{(3/2)}} x_3
\left( \frac{1}{2} - x_0\right) h(x_0)\right| \, ,
\eeq
which leads to the bounds
\beqa
\left| \frac{E_4^{\rm loop}(\pi^+ K_1^0)}{E_B} \right| &\leq&
3.4 \cdot 10^{-2} , \no \\
\left| \frac{E_4^{\rm loop}(K^+ \eta)}{E_B} \right| &\leq&
0.7 \cdot 10^{-2} .
\eeqa
At least in the foreseeable future, the loop amplitude can safely be
neglected in comparison with the bremsstrahlung amplitude (\ref{EBKP}).
On the other hand, the counterterm amplitude (\ref{E4KP})  should be
within reach of facilities with intense $K^+$ beams such as DAPHNE
\cite{DAF}, not to speak of proper kaon factories.

For $K_L \ra \pi^+ \pi^- \gamma$, it was essential to include $V$
exchange effects of $O(p^6)$, in particular to understand the slope
parameter $c$. All the mechanisms discussed there also contribute to
$K^+ \ra \pi^+ \pi^0 \gamma$. The weak VMD amplitude is \cite{ENP1}
\beq
M_6^{\rm VMD} = - C_V
\eeq
and the direct weak amplitude is given by ($M_\pi = 0$)
\beq
M_6^{\rm FM} = 2k_f C_V + \frac{e G_8 M_K^5 L_9}{2\pi^2 F^3} k_f
(3 - 8x_+ - 2x_0)
\eeq
in the framework of the FM. Altogether, we find for the total magnetic
amplitude
\beq
M = M_4 + M_6 = \frac{e G_8 M_K^3}{4 \pi^2 F} \left\{ -2 +3a_2 - 6a_3
+ r_V(2k_f -1) + \frac{2L_9 M_K^2}{F^2} k_f (3-8x_+ -2x_0)\right\}.
\label{MKP}
\eeq
Under the assumption that direct emission is entirely due to the magnetic
part, experiments \cite{KPexp,RPP} find a branching ratio
\beq
BR(55 < T_{\pi^+}\mbox{(MeV)} < 90) = (1.8 \pm 0.4) \cdot 10^{-5}
\eeq
for the given cuts in the kinetic energy of the charged pion.
Proceeding in a similar way as for $K_L \ra \pi^+ \pi^- \gamma$, we
extract the quantity $A_4 = -2 + 3a_2 - 6a_3$ from the measured rate.
In order to exhibit the sensitivity to the $O(p^6)$ contributions, we
first determine $A_4$ in the limit where $V$ exchange is turned off
($r_V = L_9 = 0$):
\beq
A_4 = - 4.5 \pm 0.5.
\eeq
For the physical values of $r_V$ and $L_9$ listed in Eq. (\ref{MKL}),
$A_4$ is found to be
\beq
A_4 = - 4.1 - 0.3 k_f \pm 0.5, \qquad 0 \leq k_f \leq 1.
\eeq

\noindent We draw the following conclusions:
\begin{enumerate}
\item[i.] Compared with $K_L \ra \pi^+ \pi^- \gamma$, the $V$ exchange
contributions are of less importance in the present case. Especially
for $k_f \simeq 1$, the $O(p^6)$ terms are essentially negligible in the
rate.
\item[ii.] However, the last term in Eq. (\ref{MKP}) shows a rather
pronounced dependence on $x_+$. A high-precision analysis of the decay
distribution in $T_{\pi^+}$ may be able to reveal this dependence.
\item[iii.] The fitted values of $A_4$ are very much consistent with our
expectations based on $a_i \; \letsmod \; 1$ (cf. Sect.~3).
\item[iv.] Because of the expected positive interference between $E_B$
and $E_B^{\rm local}$, the coefficient $|A_4|$ is probably somewhat
smaller than found above. Future experimental analysis should include
an E1 amplitude of the type (\ref{E4KP}).
\end{enumerate}

\renewcommand{\theequation}{\arabic{section}.\arabic{equation}}
\setcounter{equation}{0}
\section{Survey of the decays $K \ra \pi \pi \gamma \gamma$ and
$K \ra \pi \pi \pi \gamma (\gamma)$}
The complete list of non-leptonic $K$ decays with direct anomalous
contributions can be found in Table \ref{taban}. In comparison with the
dominant decays $K_L \ra \pi^+ \pi^- \gamma$, $K^+ \ra \pi^+ \pi^0
\gamma$ discussed in the previous section, the remaining processes are
either suppressed by phase space or by the presence of an extra photon
in the final state. It seems premature to perform a complete analysis
of all those transitions to $O(p^4)$ in CHPT. Instead, we discuss their
general features and illustrate the expected magnitude of anomalous
contributions for two specific examples.

For the decays $K \ra \pi\pi\gamma\gamma$, the general theorem of Sect.~4
applies. Thus, the $O(p^2)$ amplitude is completely given by
bremsstrahlung. As for the $K \ra \pi\pi\gamma$ transitions, direct
anomalous amplitudes occur again only in the decays $K^+ \ra \pi^+ \pi^0
\gamma\gamma$ and $K_L \ra \pi^+ \pi^- \gamma\gamma$ where bremsstrahlung
is suppressed. However, in both cases the dominant anomalous contributions
are not the direct ones, but the rather trivial $\pi^0 \ra 2\gamma$
transitions from $K \ra 3\pi$ intermediate states.\footnote{Also
$\eta \ra 2\gamma$ contributes, but to a much lesser extent.} Therefore,
in order to isolate the non-trivial anomalous amplitudes in both
$K^+ \ra \pi^+ \pi^0 \gamma\gamma$ and $K_L \ra \pi^+ \pi^-\gamma\gamma$
it is necessary to stay away from the pion pole in the $2\gamma$-invariant
mass. In practice, only the part of phase space with large $m_{2\gamma}$
seems feasible for this purpose.

\goodbreak

Let us consider the decay $K^+(P)\ra \pi^+ (p_+) \pi^0(p_0)
\gamma(q_1) \gamma(q_2)$ as an example. As explained above,
the amplitude at $O(p^2)$ is completely determined by
$A(K^+ \ra \pi^+ \pi^0)$, which is suppressed by the $\Delta I = 1/2$
rule:
\beqa
\lefteqn{E^{(1)} \ve^\mu(q_1)^* \ve^\nu(q_2)^* \cdot } \no \\
&& \cdot \left\{ - \frac{1}{Pq_1} \frac{1}{p_+ q_2} P_\mu p_{+\nu}
- \frac{1}{Pq_2} \frac{1}{p_+ q_1} p_{+\mu} P_\nu + \right. \no \\
&& \mbox{} + \frac{1}{P(q_1 + q_2) - q_1 q_2} \left[ \frac{1}{Pq_1}
P_\mu (P-q_1)_\nu + \frac{1}{Pq_2} (P - q_2)_\mu P_\nu + g_{\mu\nu} \right]
\no \\
&& \mbox{} + \left.\frac{1}{p_+ (q_1 + q_2) + q_1 q_2} \left[
\frac{1}{p_+ q_1} p_{+\mu} (p_+ + q_1)_\nu + \frac{1}{p_+ q_2}
(p_+ + q_2)_\mu p_{+\nu} - g_{\mu\nu} \right] \right\},
\label{E1}
\eeqa
with $E^{(1)} = e^2 A(K^+ \ra \pi^+ \pi^0)$.

At $O(p^4)$, the electric amplitude is given by
\beqa
\lefteqn{ \ve^\mu(q_1)^* \ve^\nu(q_2)^* \cdot } \no \\
&& \cdot \left\{E^{(2)} \left[ \frac{1}{Pq_1} P_\mu
(p_+ q_2 p_{0\nu} - p_0 q_2 p_{+\nu}) +
\frac{1}{Pq_2} (p_+ q_1 p_{0\mu} - p_0 q_1 p_{+\mu})
P_\nu  \right.\right. \no \\
&& \mbox{} - \frac{1}{p_+q_1} p_{+\mu}((p_+ + q_1)q_2 p_{0\nu} - p_0 q_2
(p_+ + q_1)_\nu)  \no \\
&& \mbox{} - \frac{1}{p_+ q_2} ((p_+ + q_2)q_1 p_{0\mu} - p_0 q_1
(p_+ + q_2)_\mu) p_{+\nu} \no \\
&& \Biggl.\Biggl.\mbox{} + (p_{0\mu} q_{1\nu} + q_{2\mu} p_{0\nu} - (q_1 + q_2)
p_0 g_{\mu\nu}) \Biggr] +
E^{(3)}(q_{2\mu} q_{1\nu} - q_1 q_2 g_{\mu\nu}) \Biggr\},
\label{E23}
\eeqa
where the coefficients $E^{(2)}$ and $E^{(3)}$ are scale-independent
combinations of the coupling constants $N_i$:
\beqa
E^{(2)} &=& \frac{8 \pi i\alpha G_8}{F} (N_{14} - N_{15} - N_{16} -
N_{17}), \no \\
E^{(3)} &=& \frac{32 \pi i\alpha G_8}{3F} (N_{14} - N_{15} - 2N_{18}).
\label{E2E3}
\eeqa
Consequently, the one-loop contributions to the decay amplitude must be
finite. From the similarity with the case of $K^+ \ra \pi^+ \pi^0 \gamma$,
we expect them to be small.

The contributions from the chiral anomaly enter in the magnetic amplitude,
which is given by
\beqa
\lefteqn{\ve^\mu(q_1)^* \ve^\nu(q_2)^* \ve^{\alpha\beta\gamma\delta}
\cdot } \no \\
&& \cdot \left\{ M^{(1)} \left[ - g_{\mu\delta} p_{+\gamma} q_{1\alpha}
p_{0\beta} \frac{P_\nu}{Pq_2} - g_{\nu\delta} p_{+\gamma}q_{2\alpha}
p_{0\beta} \frac{P_\mu}{Pq_1}  \right.\right. \no \\
&& \left.\mbox{} + g_{\mu \delta}(p_+ + q_2)_\gamma q_{1\alpha} p_{0\beta}
\frac{p_{+\nu}}{p_+q_2} + g_{\nu\delta}(p_+ + q_1)_\gamma q_{2\alpha}
p_{0\beta} \frac{p_{+\mu}}{p_+ q_1} + g_{\mu\gamma} g_{\nu\delta}
(q_1 - q_2)_\alpha p_{0\beta}\right]  \no \\
&& \mbox{} + \Biggl. (M^{(2)} + M^{(3)} + M^{(4)}) g_{\mu\gamma}
g_{\nu\delta} q_{1\alpha} q_{2\beta}\Biggr\},
\label{MAMP}
\eeqa
with
\beqa
M^{(1)} &=& \frac{2\alpha G_8}{\pi F} \left(1 - \frac{3}{2} a_2 + 3a_3\right),
\no \\
M^{(2)} &=& \frac{4\alpha G_8}{3\pi F} ,\no \\
M^{(3)} &=& \frac{\alpha G_8}{3\pi F}
\frac{6 Pp_+ - 3M_K^2 - 2M_\pi^2 + 2(q_1 + q_2)^2}{(q_1 + q_2)^2
- M_\pi^2} ,\no \\
M^{(4)} &=& \frac{2\alpha G_8}{\pi F} \frac{P(p_+ - p_0)}{(q_1 + q_2)^2
- M_\eta^2}.
\label{M1234}
\eeqa
The term with $M^{(1)}$ is determined by the magnetic amplitude of
$K^+ \ra \pi^+  \pi^0 \gamma$. The second term, proportional to
$M^{(2)}$, is generated by the Lagrangian (3.3). Finally, the last two
terms are coming from a $K^+ \ra \pi^+ \pi^0 \pi^0$
$(K^+ \ra \pi^+ \pi^0 \eta)$ intermediate state, followed by a subsequent
transition $\pi^0 \ra \gamma\gamma$ $(\eta \ra \gamma\gamma)$.

As long as the photon helicities are not measured, there is no interference
between electric and magnetic amplitudes. Our numerical results for the
various contributions to the branching ratio with three different cuts
in the $2\gamma$ invariant mass $m^2_{2\gamma} = (q_1 + q_2)^2$ are
displayed in Tables 3 and 4. For the coupling constants $N_i$ occurring
in (\ref{E2E3}) we have chosen the values suggested by the FM with
$k_f = 1$. In the quantity $M^{(1)}$ of (\ref{M1234}) we have used
the na\"{\i}ve factorization values $a_2 = a_3 = 1$.
\begin{table}
\caption{Contributions to the branching ratio $BR(K^+ \ra \pi^+ \pi^0
\gamma \gamma)$ in units of $10^{-11}$
from the electric amplitudes. The indices $i,j = 1,2,3$
refer to $E^{(1)},E^{(2)},E^{(3)}$ of
(\protect\ref{E1}) and (\protect\ref{E23}) .
}
$$
\begin{tabular}{|c|c|c|c|} \hline
$i,j$ &
$BR_{ij}(m_{2\gamma} > 170$ MeV)  &
$BR_{ij}(m_{2\gamma} > 180$ MeV)  &
$BR_{ij}(m_{2\gamma} > 190$ MeV)    \\ \hline \hline
1,1 & 11.1 & 4.3 & 1.3 \\ \hline
2,2 & 0.7 &  0.3  & 0.1 \\ \hline
3,3 & 0.7 & 0.4 & 0.2  \\ \hline
1,2 & 5.4 & 2.3 & 0.7 \\ \hline
1,3 & 4.8 & 2.4 & 0.9 \\ \hline
2,3 & 1.3 & 0.7 & 0.3 \\ \hline \hline
sum & 23.9 & 10.3 & 3.5 \\ \hline
\end{tabular}
$$
\end{table}
\begin{table}
\caption{Contributions to the branching ratio $BR(K^+ \ra \pi^+ \pi^0
\gamma \gamma)$ in units of $10^{-11}$
from the magnetic amplitudes. The indices $i,j = 1,2,3,4$
refer to $M^{(1)},M^{(2)},M^{(3)},M^{(4)}$ of
(\protect\ref{MAMP}).
}
$$
\begin{tabular}{|c|c|c|c|} \hline
$i,j$ &
$BR_{ij}(m_{2\gamma} > 170$ MeV)  &
$BR_{ij}(m_{2\gamma} > 180$ MeV)  &
$BR_{ij}(m_{2\gamma} > 190$ MeV)   \\ \hline \hline
1,1 & 47.2 & 22.2 & 8.0 \\ \hline
2,2 & 0.9 &  0.5  & 0.2 \\ \hline
3,3 & 22.0 & 8.5 & 2.7  \\ \hline
4,4 & 0.0 & 0.0 & 0.0 \\ \hline
1,2 & 12.7 & 6.7 & 2.7 \\ \hline
1,3 & -- 61.7 & -- 26.9 & -- 9.2 \\ \hline
1,4 & -- 1.9 & -- 0.9 & -- 0.3 \\ \hline
2,3 & -- 4.4 & -- 2.1 & -- 0.8 \\ \hline
2,4 & -- 0.1 & -- 0.1 & 0.0 \\ \hline
3,4 & 0.7 & 0.3 & 0.1 \\ \hline \hline
sum & 15.2 & 8.3 & 3.4 \\ \hline
\end{tabular}
$$
\end{table}

The remaining cases $K \ra 3\pi\gamma(\gamma)$ are in general dominated
by the bremsstrahlung amplitudes entering with full strength. However, it
is important to realize that the theorem of Sect.~4 is not applicable
for those decays. In other words, already at $O(p^2)$ in CHPT there are
in general additional contributions to the amplitudes that are not of
the bremsstrahlung type. Those leading-order contributions are interesting
in themselves and will be discussed in detail elsewhere. Here, we are
interested in the sensitivity required to detect the presence of direct
anomalous amplitudes governed by the weak coupling constants $N_{28},
\ldots,N_{31}$. We emphasize that there are in addition anomalous
amplitudes of the reducible type that are not covered by the Lagrangian
(\ref{Law}), but are of the type shown in Fig. \ref{redaa}. Since
the WZW functional has no free parameters, those contributions are
completely determined by the octet coupling $G_8$ in the $O(p^2)$ weak
Lagrangian (\ref{L2w}). An example, in addition to the
$\pi^0(\eta) \ra \gamma\gamma$ vertex relevant for two-photon decays, is
a transition of the type $K \ra \pi\pi \ra \pi\pi\pi\gamma$ where the
second step occurs via the WZW functional.

An interesting case with direct anomalous contributions is provided by
$K_L \ra \pi^+ \pi^- \pi^0 \gamma$. From the explicit form of the octet
operators $W_{28},\ldots,W_{31}$ in Eq. (\ref{WiK}) and using
(\ref{Nan}), one finds the direct anomalous coupling
\beq
\cL_{\rm anom}^{\rm direct} (K_L \ra \pi^+ \pi^- \pi^0 \gamma) =
\frac{3 e G_8}{16\pi^2 F^2} (8a_1 + a_2 - 10a_3) \wt F^{\mu\nu}
\partial_\mu K_2^0 \partial_\nu \pi^0 \pi^+ \pi^-.
\label{KL3}
\eeq
This Lagrangian has a few interesting features. First of all, the
coefficients are potentially rather big if we recall $a_i = O(1)$ from
the dominance of factorizable contributions (Sect.~3). However, with
the na\"{\i}ve factorization values $a_i = 1$ there would be almost complete
destructive interference in (\ref{KL3}). Finally, the decay
$K_L \ra \pi^+ \pi^- \pi^0 \gamma$ is the only experimentally accessible
process sensitive to the weak coupling constant $N_{28}$. Therefore,
this decay affords in principle an interesting possibility to check the
structure of direct anomalous terms.

Unfortunately, the available phase space is small. Ignoring all other
contributions, in particular the dominant bremsstrahlung amplitude, the
Lagrangian (\ref{KL3}) would give rise to a branching ratio
\beq
\left.BR(K_L \ra \pi^+ \pi^- \pi^0 \gamma)\right|_{\rm direct} =
(8a_1 + a_2 - 10a_3)^2 \cdot 2 \cdot 10^{-10}.
\eeq
Since not even the bremsstrahlung part has so far been observed
experimentally, a test of the anomalous coupling (\ref{KL3}) may have
to wait for a while.

\renewcommand{\theequation}{\arabic{section}.\arabic{equation}}
\setcounter{equation}{0}
\section{Summary}
Anomalies play a fundamental role in our understanding of modern particle
physics. Gauge symmetries should be free of anomalies in order
to allow a consistent quantization of the corresponding field theory.
Global symmetries, however, can be broken at the quantum level.
A well-known example is the chiral anomaly \cite{ABJ,Bard},
present in quantum field
theories with chiral structure, such as  the standard model.
They do not constitute any obstruction for a proper quantization.
Moreover, they have important implications for particle physics.

In the standard model, the chiral anomaly manifests itself most directly
in the low-energy interactions of the pseudo-Goldstone bosons
of spontaneously broken chiral symmetry.
Since anomalies have a short-distance origin, their effect is completely
calculable. The translation from the fundamental quark-gluon
level to the effective chiral level (mesons) is unaffected by
hadronization problems.
The Wess-Zumino-Witten functional \cite{WZW} encodes all low-energy
manifestations of the chiral anomaly in strong interactions, in the presence
of arbitrary external vector and axial-vector fields.

It is straightforward to work out the experimental
consequences of the anomaly for electromagnetic and semileptonic weak
processes.
In addition to the classical test via the two-photon decays of the
neutral pseudoscalars ($\pi^0\to\gamma\gamma$, $\eta\to\gamma\gamma$,
$\eta'\to\gamma\gamma$)
or the $\gamma 3\pi$ and $\gamma\pi^+\pi^-\eta$ interactions \cite{BijRev},
the manifestations of the non-Abelian chiral anomaly have
mainly been investigated
in semileptonic kaon decays \cite{Ksemi,EckMor,old-tests}.
Tau decays into three or more hadrons have also been pointed out
\cite{kramer,pich,ggp} to be sensitive to the anomaly, especially the
decays $\tau\to\nu_\tau\eta + n \pi$ ($n\ge 2$) \cite{kramer,pich},
which (for small hadronic invariant mass) only get contributions
from the WZW term. Unfortunately, the presence of resonances at the high
$Q^2$ values relevant for the $\tau$ decay spoils the possibility of
making a clean quantitative test of the anomaly predictions.

In this paper we have
presented a systematic investigation of the relevance of the chiral anomaly
in non-leptonic weak transitions.
Within the framework of CHPT, the manifestations of the anomaly appear first
at $O(p^4)$.
They can be grouped in two different classes of anomalous amplitudes:
reducible and direct contributions.

The reducible amplitudes arise from the contraction of meson lines
between a weak $\Delta S=1$ vertex and the WZW functional
(Fig. \ref{redaa}).
The so-called pole contributions can be given in closed form \cite{ENP1}
as a local Lagrangian (\ref{Law}) which contributes only to the decays
$K^+\to\pi^+\pi^0\gamma$,  $K^+\to\pi^+\pi^0\gamma\gamma$ and
$K_L\to\pi^+\pi^-\gamma\gamma$.
There are other reducible contributions which cannot be written in
local form.
In the octet limit, all reducible anomalous amplitudes of $O(p^4)$ can be
predicted in terms of the coupling $G_8$.

The direct anomalous contributions arise from the contraction of the
W boson field between a strong Green function on one side and the WZW
functional on the other side.
Their computation is not straightforward,
because of the presence of strongly interacting fields on both sides
of the $W$.
Using the operator product expansion to integrate out the heavy fields
($W$, $t$, $b$, $c$), one gets an effective Hamiltonian in terms of
four-quark operators, which must be realized
at the bosonic level in the presence of the anomaly.
The factorizable
contribution can be calculated in terms of bosonic currents.
Due to the non-renormalization theorem of the chiral anomaly \cite{AB},
there are no QCD corrections to the anomalous current, which is directly
obtained from the WZW functional.
Moreover, the non-factorizable piece does not get any contribution from the WZW
functional. Therefore, the bosonized form of the direct anomalous amplitude
can be fully predicted \cite{BEP1}.
At $O(p^4)$, the anomaly turns out to contribute to all the possible
octet operators proportional to the $\varepsilon$ tensor
($W_{28}$, $W_{29}$, $W_{30}$ and $W_{31}$).
In spite of its anomalous origin, this contribution is
chiral-invariant.
Unfortunately, the coefficients of these four operators
get also non-factorizable contributions of non-anomalous origin,
which cannot be computed in a model-independent way.
Therefore, we can only parametrize the final result [Eq. (\ref{Nan})]
in terms of dimensionless coefficients $a_i$ ($i=1,\ldots,4$), which are
expected to be positive and of order one.

  A complete list of all kinematically allowed non-leptonic $K$ decays
that get local contributions from the anomaly
at $O(p^4)$ is given in Table \ref{taban}.
Only radiative $K$ decays are sensitive to the anomaly in the
non-leptonic sector.
The most frequent ``anomalous'' decays $K_L\to\pi^+\pi^-\gamma$ and
$K^+\to\pi^+\pi^0\gamma$ share the remarkable feature that the
normally dominant bremsstrahlung amplitude is strongly suppressed,
making the experimental verification of the anomalous amplitude
substantially easier.
This suppression has different origins: $K^+\to\pi^+\pi^0$ proceeds
through the small 27-plet part of the non-leptonic weak interactions,
whereas $K_L\to\pi^+\pi^-$ is CP-violating.

For $K_L\to\pi^+\pi^-\gamma$, the direct emission rate is completely
dominated by the magnetic amplitude.
There is however a strong destructive interference between the
$O(p^4)$ contribution (\ref{KLM4})
and the anomalous reducible amplitude (\ref{M6an}),
first appearing at $O(p^6)$.
This $O(p^6)$ contribution stems from corrections to the
Gell-Mann-Okubo mass formula and is very sensitive to the $\eta$--$\eta'$
mixing angle and to nonet-symmetry-breaking effects.
This makes a reliable estimate of the rate very difficult.
Moreover, there is an important
VMD contribution at $O(p^6)$, which generates a sizeable dependence of
the magnetic amplitude on the photon energy.
Although we cannot make absolute predictions for this decay,
CHPT establishes a correlation between the rate and the energy slope
in agreement with experiment.

 For $K^+\to\pi^+\pi^0\gamma$, there is a potentially sizeable
electric amplitude interfering with bremsstrahlung.
This interference must be taken into account in the experimental analysis
to extract the contribution of the anomaly to the rate.
The VMD contribution to the magnetic amplitude is less important than in
$K_L\to\pi^+\pi^-\gamma$. Nevertheless, it generates a rather pronounced
dependence on the charged pion energy.
Fitting our formulae to the measured direct-emission rate, one gets a
value for the anomalous $O(p^4)$ magnetic amplitude (\ref{M4Kp}),
in good agreement with the factorization estimate.

The remaining non-leptonic $K$ decays with direct anomalous contributions
(Table \ref{taban}) are either suppressed by phase space or by the presence
of an
extra photon in the final state. For the decays $K\to\pi\pi\gamma\gamma$,
direct anomalous amplitudes occur again only for
$K_L\to\pi^+\pi^-\gamma\gamma$ and $K^+\to\pi^+\pi^0\gamma\gamma$,
where bremsstrahlung is suppressed. However, in both cases the dominant
anomalous contributions are the trivial $\pi^0\to 2\gamma$ transitions
from $K\to 3\pi$ intermediate states.
The decays $K\to 3\pi\gamma(\gamma)$ are in general dominated by the
bremsstrahlung amplitudes entering with full strength.
An interesting case with direct anomalous contributions is provided by
$K_L\to\pi^+\pi^-\pi^0\gamma$, which is the only experimentally
accessible process sensitive to the weak coupling $N_{28}$.

Although not as straightforward as for electromagnetic and semileptonic
weak processes, non-leptonic $K$ decays offer interesting possibilities
for experimental tests of the chiral anomaly.

\vspace{1cm}
\noindent {\bf Acknowledgements}

\noindent We are grateful to Erik Ramberg for information on the
results of Ref. \cite{E731}.

\newpage

\appendix{\section*{Appendix A: Bremsstrahlung amplitudes for \\
\mbox{} \hspace*{3.7cm} $K \ra \pi \pi \gamma \ldots \gamma$}}
\newcounter{zahler}
\renewcommand{\thesection}{\Alph{zahler}}
\renewcommand{\theequation}{\Alph{zahler}.\arabic{equation}}
\setcounter{zahler}{1}
\setcounter{equation}{0}

The tree level generating functional $Z_{\rm tree}[j,A]$ for connected
$(3+n)$-point functions with three external spin-0 legs and $n$ external
photons can be written as
\beq
Z_{\rm tree}[j,A] = \int d^4x \cL_2^{\rm cubic} (\vp_k^{\rm cl}[j,A],
D_\mu \vp_k^{\rm cl}[j,A]). \label{Ztree}
\eeq
$\cL_2^{\rm cubic}$ is the cubic part of the general Lagrangian
$\cL_2(\vp_k,D_\mu \vp_k)$ in the theorem of Sect. 4. The classical
fields $\vp_k[j,A]$ are solutions of the free equations of motion
\beq
(D^2 + M_k^2)\vp_k^{\rm cl} = j_k, \qquad
D_\mu \vp_k = (\partial_\mu - i eq_k A_\mu)\vp_k, \qquad
k = +,-,0 \label{EOM}
\eeq
with external sources $j_k$ in the presence of an external electromagnetic
field $A_\mu$.

Using partial integration in the action, the most general gauge-invariant
cubic interaction Lagrangian with at most two derivatives has the form
\beq
\cL_2^{\rm cubic} = \vp_0(a_1 D^2\vp_+ \vp_- + a_2 \vp_+ D^2\vp_-
+ a_3 D_\mu \vp_+ D^\mu \vp_-) + b \vp_0 \vp_+ \vp_-,
\eeq
where $a_1,a_2,a_3,b$ are coupling constants that may depend on the
masses $M_k$. We use partial integration once more,
\beq
\vp_0 D_\mu \vp_+ D^\mu \vp_- \; \wh{=} \; \frac{1}{2}
(D^2 \vp_0 \vp_+ \vp_- - \vp_0 D^2 \vp_+ \vp_- - \vp_0 \vp_+ D^2\vp_-),
\eeq
to bring $\cL_2^{\rm cubic}$ into the final form
\beq
\cL_2^{\rm cubic} = \vp_0(a'_1 D^2 \vp_+\vp_- + a'_2 \vp_+ D^2 \vp_-)
+ a'_3 D^2 \vp_0 \vp_+ \vp_- + b \vp_0 \vp_+ \vp_-.
\eeq

In the generating functional $Z_{\rm tree}$ in (\ref{Ztree}), the
derivatives appear therefore only in the form of covariant d'Alembertians
acting on the classical fields $\vp_k^{\rm cl}$. Using the equations
of motion (\ref{EOM}), we may write
\beq
D^2 \vp_k^{\rm cl} = D^2 (D^2 + M_k^2)^{-1} j_k = j_k - M_k^2
(D^2 + M_k^2)^{-1} j_k.
\eeq
Since the first term $j_k$ on the right-hand side does not contribute
to on-shell amplitudes (amputated Green functions), we may replace
$D^2 \vp_k^{\rm cl}$ by $- M_k^2 \vp_k^{\rm cl}$ everywhere in
$\cL_2^{\rm cubic}(\vp_k^{\rm cl},D_\mu \vp_k^{\rm cl})$. Thus, the
Lagrangian is equivalent to the non-derivative cubic Lagrangian
\beq
\cL_2^{\rm cubic} \; \wh = \; \vp_0 \vp_+ \vp_-
(- a'_1 M_+^2 - a'_2 M_-^2 - a'_3 M_0^2 + b)
\eeq
for which the theorem is trivially satisfied.

\newpage
\appendix{\section*{Appendix B: Loop functions}}
\renewcommand{\thesection}{\Alph{zahler}}
\renewcommand{\theequation}{\Alph{zahler}.\arabic{equation}}
\setcounter{zahler}{2}
\setcounter{equation}{0}

The kinematical functions $g(x)$ and $h(x)$ originate from the loop
integral
\beq
\int \frac{d^4k}{(2\pi)^4} \dfrac{k^\mu k^\nu}{(k^2 - M_1^2)
((k+q)^2 - M_1^2)((k-p)^2 - M_2^2)}  =
i g^{\mu\nu} C_{20}(p^2,(p+q)^2,M_1^2,M_2^2) + \ldots,
\label{B1}
\eeq
where $q^2 = 0$. The dots in (\ref{B1}) refer to terms that are irrelevant
in our case. In the next step we define the (finite) function
\beq
\ol{C}_{20}(p^2,(p+q)^2,M_1^2,M_2^2) = C_{20}(p^2,(p+q)^2,M_1^2,M_2^2)
- C_{20}(p^2,p^2,M_1^2,M_2^2).
\label{B2}
\eeq
With this definition, the function $g(x)$ [used in (\ref{gx})] and the
function $h(x)$ of (\ref{H}) are given by
\beqa
g(x) &=& (4\pi)^2 \frac{M_K^2}{pq} \left[\ol{C}_{20}(p^2,(p+q)^2,M_\pi^2,
M_K^2) + \ol{C}_{20}(p^2,(p+q)^2,M_K^2,M_\eta^2)\right],
\quad
\no \\
h(x) &=& (4\pi)^2 \frac{M_K^2}{pq} \left[\ol{C}_{20}(p^2,(p+q)^2,M_\pi^2,
M_K^2) + \frac{2}{3} \ol{C}_{20}(p^2,(p+q)^2,M_K^2,M_\eta^2)\right],
\quad
\eeqa
where $p^2 = M_\pi^2$ and $pq = M_K^2(\frac{1}{2} - x)$.

\newpage
\newcommand{\PL}[3]{{Phys. Lett.}        {#1} {(19#2)} {#3}}
\newcommand{\PRL}[3]{{Phys. Rev. Lett.} {#1} {(19#2)} {#3}}
\newcommand{\PR}[3]{{Phys. Rev.}        {#1} {(19#2)} {#3}}
\newcommand{\NP}[3]{{Nucl. Phys.}        {#1} {(19#2)} {#3}}

\end{document}